\documentclass[journal]{IEEEtran}
\usepackage{amsmath,amsfonts}
\usepackage{algorithmic}
\usepackage{array}
\usepackage[caption=false,font=normalsize,labelfont=sf,textfont=sf]{subfig}
\usepackage{textcomp}
\usepackage{stfloats}
\usepackage{url}
\usepackage{verbatim}
\usepackage{graphicx}
\usepackage{color}
\usepackage{pifont}
\usepackage[table,xcdraw]{xcolor}
\usepackage{threeparttable}
\usepackage{makecell}
\usepackage{multirow}
\usepackage{booktabs}
\usepackage{cite}
\usepackage[linesnumbered,ruled,vlined]{algorithm2e}
\usepackage{amssymb} 
\usepackage{hyperref}
\usepackage{float}
\usepackage{listings}
\makeatletter
\def\UrlAlphabet{%
      \do\a\do\b\do\c\do\d\do\e\do\f\do\g\do\h\do\i\do\j%
      \do\k\do\l\do\m\do\n\do\o\do\p\do\q\do\r\do\s\do\t%
      \do\u\do\v\do\w\do\x\do\y\do\z\do\A\do\B\do\C\do\D%
      \do\E\do\F\do\G\do\H\do\I\do\J\do\K\do\L\do\M\do\N%
      \do\O\do\P\do\Q\do\R\do\S\do\T\do\U\do\V\do\W\do\X%
      \do\Y\do\Z}
\def\UrlDigits{\do\1\do\2\do\3\do\4\do\5\do\6\do\7\do\8\do\9\do\0}
\g@addto@macro{\UrlBreaks}{\UrlOrds}
\g@addto@macro{\UrlBreaks}{\UrlAlphabet}
\g@addto@macro{\UrlBreaks}{\UrlDigits}
\makeatother
\lstdefinelanguage[RISC-V]{Assembler}
{
  alsoletter={.}, 
  morekeywords=[1]{ 
    lb, lh, lw, lbu, lhu,
    sb, sh, sw,
    sll, slli, srl, srli, sra, srai,
    add, addi, sub, lui, auipc,
    xor, xori, or, ori, and, andi,
    slt, slti, sltu, sltiu,
    beq, bne, blt, bge, bltu, bgeu,
    j, jr, jal, jalr, ret,
    scall, break, nop,
    addq, cmpq, jne, incd, whilelt, b.first, vsetvli, bnez, vsetcsr, vns.add
  },
  morekeywords=[2]{ 
    .align, .ascii, .asciiz, .byte, .data, .double, .extern,
    .float, .globl, .half, .kdata, .ktext, .set, .space, .text, .word
  },
  morekeywords=[3]{ 
    zero, ra, sp, gp, tp, s0, fp,
    t0, t1, t2, t3, t4, t5, t6,
    s1, s2, s3, s4, s5, s6, s7, s8, s9, s10, s11,
    a0, a1, a2, a3, a4, a5, a6, a7,
    ft0, ft1, ft2, ft3, ft4, ft5, ft6, ft7,
    fs0, fs1, fs2, fs3, fs4, fs5, fs6, fs7, fs8, fs9, fs10, fs11,
    fa0, fa1, fa2, fa3, fa4, fa5, fa6, fa7,
    p0.d, rax, x3, x4, vns0, vns3, vns6
  },
  morecomment=[l]{;},   
  morecomment=[l]{\#},  
  morestring=[b]",      
  morestring=[b]'       
}

\definecolor{codegreen}{rgb}{0,0.6,0}
\definecolor{codegray}{rgb}{0.5,0.5,0.5}
\definecolor{codepurple}{rgb}{0.58,0,0.82}
\definecolor{backcolour}{rgb}{0.95,0.95,0.92}
\lstdefinestyle{mystyle}{
    frame=single,
    backgroundcolor=\color{backcolour},   
    commentstyle=\color{codegreen},
    keywordstyle=[1]\color{blue!80!black},        
    keywordstyle=[2]\color{orange!80!black},      
    keywordstyle=[3]\color{red!80!black},         
    numberstyle=\tiny\color{codegray},
    stringstyle=\color{codepurple},
    basicstyle=\ttfamily\footnotesize,
    breakatwhitespace=false,         
    breaklines=true,                 
    captionpos=b,                    
    keepspaces=true,                                  
    showspaces=false,                
    showstringspaces=false,
    showtabs=false,                  
    tabsize=2
}

\begin{document}

\title{Unlimited Vector Processing for Wireless Baseband Based on RISC-V Extension}

\author{Limin Jiang, Yi Shi, Yihao Shen, Shan Cao, Zhiyuan Jiang, and Sheng Zhou
\thanks{This work was supported in part by the National Natural Science Foundation of China (NSFC) under Grants 62271300 and 12141107, in part by the Shanghai Municipal Science and Technology Commission under grants 24DP1501100 and 24DP1500600. The corresponding authors are Shan Cao and Zhiyuan Jiang.

L. Jiang, Y. Shi, Y. Shen, S. Cao, and Z. Jiang are with Shanghai Key Laboratory of Chips and Systems for Intelligent Connected Vehicle and School of Communication and Information Engineering, Shanghai University, Shanghai 200444, China. E-mails: \{jianglimin, yishi1996, shenyihao, cshan, jiangzhiyuan\}@shu.edu.cn

S. Zhou is with Beijing National Research Center for Information Science and Technology, Department of Electronic Engineering, Tsinghua University, Beijing 100084, China. E-mail: sheng.zhou@tsinghua.edu.cn}}



\maketitle

\begin{abstract}
Wireless baseband processing (WBP) serves as an ideal scenario for utilizing vector processing, which excels in managing data-parallel operations due to its parallel structure. However, conventional vector architectures face certain constraints such as limited vector register sizes, reliance on power-of-two vector length multipliers, and vector permutation capabilities tied to specific architectures. To address these challenges, we have introduced an instruction set extension (ISE) based on RISC-V known as unlimited vector processing (UVP). This extension enhances both the flexibility and efficiency of vector computations. UVP employs a novel programming model that supports non-power-of-two register groupings and hardware strip-mining, thus enabling smooth handling of vectors of varying lengths while reducing the software strip-mining burden. Vector instructions are categorized into symmetric and asymmetric classes, complemented by specialized load/store strategies to optimize execution. Moreover, we present a hardware implementation of UVP featuring sophisticated hazard detection mechanisms, optimized pipelines for symmetric tasks such as fixed-point multiplication and division, and a robust permutation engine for effective asymmetric operations. Comprehensive evaluations demonstrate that UVP significantly enhances performance, achieving up to 3.0$\times$ and 2.1$\times$ speedups in matrix multiplication and fast Fourier transform (FFT) tasks, respectively, when measured against lane-based vector architectures. Our synthesized RTL for a 16-lane configuration using SMIC 40nm technology spans 0.94 mm$^2$ and achieves an area efficiency of 21.2 GOPS/mm$^2$.
\end{abstract}


\section{Introduction}

\IEEEPARstart{V}{ector} processing has been a cornerstone of high-performance computing for decades, offering a specialized approach to handling data-parallel tasks efficiently. By employing a single-instruction-multiple-data (SIMD) technique, vector processing can dramatically improve performance in applications such as scientific computing \cite{oliker2004scientific}, image processing \cite{ferreira2023design}, and machine learning \cite{jouppi2023tpu}. These advantages are enabled by vector processors that utilize vector registers to execute operations on arrays of data in parallel. This paradigm leverages data-level parallelism to achieve higher throughput compared to scalar processing.

A notable advancement in the field is the RISC-V Vector ``V'' (RVV) extension \cite{git-rvv}, which introduces a flexible and extensible approach to vector processing. Unlike traditional fixed-width vector architectures, the RVV allows dynamic vector lengths (VLs), enabling a more scalable and adaptable execution model. One of its key innovations is register grouping (RG), also referred to as vector length multipliers (LMULs). This mechanism enables the combination of multiple vector registers to form larger logical registers, allowing efficient operations on wider vectors. Conversely, it also supports splitting a register into smaller segments for finer-grained operations on narrow data. By accommodating various data widths dynamically, register grouping ensures efficient resource utilization across diverse workloads. Combined with its extensive set of arithmetic, logical, and data manipulation instructions, the RVV represents a significant leap in flexibility and capability. Moreover, as an open standard, it fosters widespread adoption and drives innovation within the computational community.

Despite its advantages, vector processing faces several challenges that limit its efficiency and applicability. One fundamental drawback of traditional vector architectures is the size constraint of vector registers, which determines the maximum data width that can be processed in a single operation. This limitation necessitates fragmenting larger datasets into smaller segments, incurring control overhead due to additional branch instructions and loop iterations. Such control dependencies can degrade performance, especially for workloads with irregular or large data sizes. Additionally, power-of-two constraints in some architectures, including the RVV's reliance on power-of-two LMULs, can lead to under-utilization of hardware resources. Workloads requiring non-power-of-two vector lengths may experience wasted computation and energy inefficiencies, as hardware resources are left idle or misaligned with data requirements.

Another challenge lies in the implementation of versatile vector permutations, which are operations that rearrange data within vectors. Such permutations are frequently required in domains like cryptography \cite{nannipieri2021risc, kuo2022risc} and wireless signal processing \cite{gautschi2017near}. However, the ability to perform arbitrary permutations efficiently is implementation-specific, varying across architectures. This lack of uniformity creates complexity in developing portable and optimized software solutions.

Just as importantly, wireless baseband processing (WBP) demands a more domain-specific architecture \cite{jiang2025hierarchical}. The computation tasks involved in WBP can be broadly categorized into two types. On one hand, tasks such as channel equalization \cite{spencer2004zero} and demodulation \cite{tosato2002simplified} are computation-intensive, requiring successive arithmetic operations. On the other hand, tasks like rate-matching \cite{behera2020efficient} involve irregular VLs and permutations, for which mainstream architectures offer limited optimization. Moreover, WBP exhibits a high tolerance for quantization error \cite{janhunen2011fixed}, enabling signal processing with fixed-point arithmetic. This allows the large silicon area typically allocated to floating-point units to be repurposed for more domain-specific fixed-point datapaths tailored to WBP, such as complex number computations and saturation logic, which will be discussed in the following sections. The question now is how to design a reasonable instruction set extension for WBP. 

To address these challenges, we propose unlimited vector processing (UVP), a novel approach that removes the dependence on fixed vector register sizes and enables seamless handling of arbitrarily long vectors. This paradigm also standardizes vector permutations to improve portability and performance across diverse computational fields. We propose an unlimited vector programming model to RISC-V instruction set extension (ISE), \textit{Xuvp}, based on insights from software programming. The programming model is implemented at register transfer level (RTL), and we detail the corresponding hardware design.  We also analyze the bottlenecks of existing vector processing approaches using benchmark kernels commonly employed in WBP. The key contributions of our work are as follows: 
\begin{itemize}
\item \textit{Formalization of UVP:} We present the ISE encoding as well as programming model for UVP. In essence, UVP supports non-power-of-two register grouping and leverages larger physical register files to handle vectors of arbitrary length. This model reduces strip-mining instructions, improving efficiency for large data sets. To further enhance flexibility, we categorize vector instructions into two groups: symmetric instructions, where source and destination operands have the same length, and asymmetric instructions, which handle operands with differing lengths. We detail distinct load/store approaches tailored for each category and demonstrate the benefits of this model through two frequently used computational kernels.
\item \textit{Hardware implementation of UVP:} We design hardware to support the UVP programming model with a new hazard detection logic optimized for the register grouping strategy. For symmetric instructions, we enhance the data pipeline to efficiently handle complex multiplication and division for fixed-point computation. For asymmetric instructions, we introduce a universal permutation engine capable of shuffling any elements to any position, accommodating arbitrary vector lengths. This hardware design ensures seamless execution of the UVP model, maximizing performance and resource utilization.
\item \textit{Evaluation of UVP:} We comprehensively evaluate UVP to demonstrate its performance and efficiency. First, we compare two frequently used kernels with a lane-based vector processing architecture, achieving up to 3.0$\times$ and 2.1$\times$ performance improvements, respectively. Next, we synthesize and implement the UVP RTL design and compare it with state-of-the-art architectures, showing superior normalized integer area efficiency. We also analyze UVP performance across different configurations, highlighting its adaptability to varying workloads. 
\end{itemize}

The remainder of this paper is organized as follows. Section \ref{section:Related} reviews and categorizes existing vector processing methodologies, introducing the fundamental strip-mining approach across different instruction set architectures (ISAs). In Section \ref{UVP}, we introduce the design principles of UVP. Section \ref{section:Hardware} details the implementation techniques used to address vector length constraints and permutation challenges. In Section \ref{section:Experiment}, we evaluate the performance of the proposed system through benchmarks and case studies. Finally, Section \ref{section:Conclusion} summarizes the conclusions of the paper.

\section{Related Work and Background}
\label{section:Related}
\subsection{Existing Vector Processing Techniques}
\subsubsection{RVV-based Vector Processors}
A significant body of work focuses on RVV-based vector processors. Both industrial and academic designs have adopted RVV for various applications, from embedded systems to high-performance computing, due to its simplicity and open-source nature. In industry, T-Head and SiFive have developed their own RVV-compatible cores, named XuanTie 910 \cite{chen2020xuantie} and X280 \cite{si5x280}, respectively. Both vector function units are supported by Linux-ready out-of-order cores. However, at the micro-architecture level, XuanTie 910 \cite{chen2020xuantie} integrates the vector function unit directly into the core pipeline, while X280 \cite{si5x280} uses a rather decoupled approach with a dedicated private interface. Additionally, AndesCore supports additional data formats beyond the RVV-specified ones \cite{nx27v}, and Semidynamics introduces a fully customizable vector processing unit \cite{semidynamic}.

In academia, the Ara vector processor continues to evolve in alignment with the RVV specification \cite{cavalcante2019ara, perotti2022new, perotti2024ara2} and is increasingly becoming a blueprint for other works \cite{humblet2024msparq, wang202430, minervini2023vitruvius+}. Data-level parallelism is achieved through scalable \textit{lanes}, with the register files divided into banks to acquire higher frequencies and reduce read/write conflicts. In contrast to lane-based designs, other RVV-compliant works implement monolithic register files, allowing each execution unit to access all vector elements within the register files \cite{platzer2021vicuna, assir2021arrow}. However, while the ISA strives to address diverse domain requirements, it may fall short in specific scenarios such as time-critical processing in wireless baseband and handling massive data loads in AI applications.

\subsubsection{Customizations for Domain-Specific Computations}
In this approach, researchers define specialized instructions on the RISC-V Base ISA, extending it to meet the needs of specific computation tasks. This customization allows for highly efficient hardware implementations tailored to particular workloads. Non-standardized vector processing accelerates data processing in specific domains by customizing critical instructions and implementing dedicated hardware logic for domain-specific tasks \cite{lee201445nm, schmidt2021eight}. In edge artificial intelligence (AI) applications, processors are often small to conserve power, and models for inference use low-precision data formats to reduce memory usage. Under these constraints, vector processing in edge AI processors is typically designed in a packed-SIMD style, with additional arithmetic logic units (ALUs) and control logic that enable parallel computation of low-precision operands within a scalar register \cite{johns2020minimal, ottavi2020mixed}. For more demanding applications, such as convolutional neural networks (CNNs), additional execution units, like tensor units or systolic arrays, are incorporated to boost throughput \cite{wang2024speed, li2022precision}. Similarly, in wireless communications, the complexity of massive-input massive-output (MIMO) systems scales exponentially with the number of antennas, leading to the development of vector processors specialized for matrix decomposition \cite{attari2022application}.

Apart from the need at the ISA-level, customization can also be applied at the micro-architectural level. Both RISC-V$^2$ \cite{patsidis2020risc} and Lazo et al. \cite{lazo2022adaptable} emphasize register file organization, implementing dedicated hardware for remapping between the logical vector register allocated by the compiler and the physical register specified by the hardware implementation. While this alleviates the complexity of register allocation algorithms for the compiler, it incurs not only increased chip area but also additional execution clock cycles per instruction at the hardware level. Furthermore, UVE \cite{domingos2021unlimited, fernandes2024functional} presents data streaming support and an emulation platform. These modifications are based on existing scalar core architectures to support various memory access patterns. Nonetheless, data streaming with indirect patterns suffers from complicated descriptors and limited parallelism.

\subsubsection{Leveraging Scalar Cores for Vector Processing}
In contrast to dedicated vector processors, some research has focused on modifying commercial-off-the-shelf scalar cores to function as vector processors. Multiple scalar cores are used in parallel, where each core processes a part of the vector in lockstep, with the instruction fetch pipeline of most cores disabled.
Essentially, the scalar cores operate together to mimic vector processing by distributing vector elements across the cores and executing the same instruction on multiple data elements simultaneously. This approach is often considered an extension or additional feature of manycore architectures \cite{ta2022big, bedoukian2021software, ottavi2023dustin}, where a set of scalar cores is utilized to handle vector workloads, rather than relying on specialized vector units.

While this approach can leverage existing scalar cores to perform vector-like operations, it tends to have lower computational efficiency compared to dedicated vector processors. This is because, although the scalar cores are working in parallel, the underlying hardware was not initially designed to handle vectorized tasks. As a result, there is often significant overhead from logic that is only useful for scalar processing, such as complex control logic and instruction fetch mechanisms that need to be reconfigured for lockstep execution. Furthermore, this approach lacks the dedicated vector units and optimizations found in true vector processors, leading to suboptimal performance in terms of throughput and energy efficiency.

\subsubsection{Digital Signal Processors for WBP}
Very long instruction word (VLIW) architectures have been widely adopted in digital signal processors (DSP) for their ability to exploit instruction-level parallelism (ILP) through static scheduling. In VLIW processors, multiple operations -- including scalar, vector, and even tensor \cite{mahurin2023qualocmm} instructions -- are bundled and issued simultaneously in a single long instruction word by the compiler \cite{lin2007palf, kuan2012compiler}. The hardware is greatly simplified compared to superscalar or vector designs, since scheduling, hazard avoidance, and resource allocation are all handled by the compiler. This makes VLIW especially attractive in domain-specific scenarios with predictable workloads and tight power constraints, namely computer vision tasks \cite{yu2025optimizing} and WBP \cite{codrescu2014hexagon, ceva2020}.

While VLIW DSPs offer high efficiency under regular and static analyzable workloads, they suffer from several limitations. First, the burden of complex scheduling and register allocation falls entirely on the compiler, which demands deep expertise in the target architecture to achieve optimal performance -- thereby reducing portability and increasing development effort. Second, static instruction bundling lacks flexibility in handling dynamic workloads or irregular data patterns, often resulting in suboptimal hardware utilization.

\subsection{Vector Strip-mining}
\label{Motivations}
\begin{figure}[!t]
  \centering
  \includegraphics[width=\linewidth]{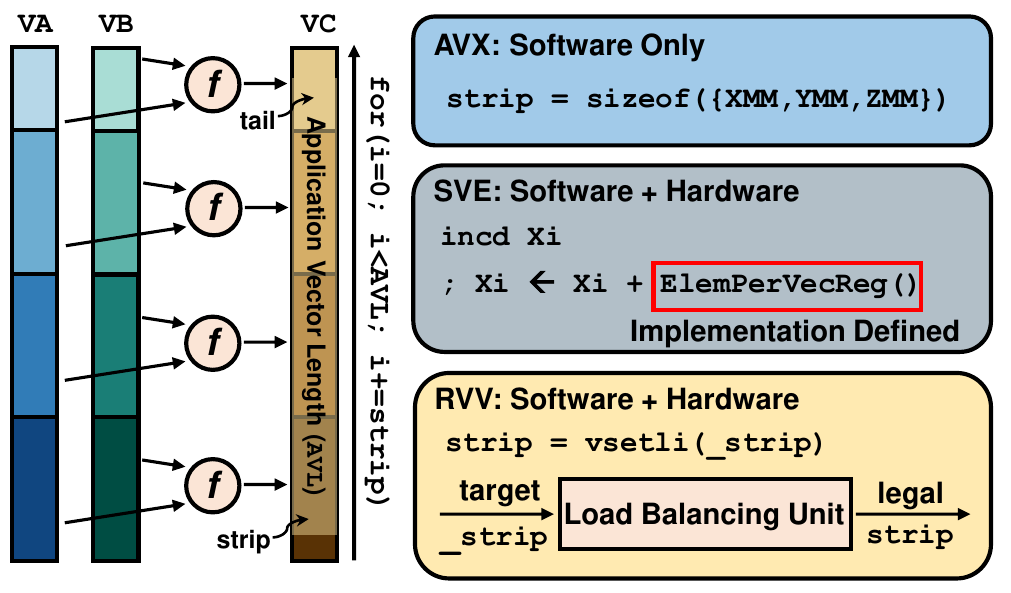}
  \vspace{-0.6cm}
  \caption{The concept of strip-mining and its implementation in mainstream ISAs.}
  \vspace{-0.6cm}
  \label{fig:strip}
\end{figure}

One of the key techniques in vector processing is strip-mining, which is used to handle data exceeding the capacity of a vector register. Fig. \ref{fig:strip} illustrates the concept of strip-mining. Essentially, strip-mining divides the data into smaller \textit{strips} that fit within the available vector register length. These strips are processed sequentially in a loop until all the data is covered. For data sizes that are not a multiple of the vector register length, additional \textit{tail} processing is required. Given two source vectors, \texttt{VA} and \texttt{VB}, with a vector length of \texttt{AVL}, a complete strip-mining procedure for an operation $f$ to compute \texttt{VC} involves first calculating $T = \lfloor \frac{AVL}{strip} \rfloor$ iterations of parallel computation fully utilizing the hardware. Subsequently, an additional computation is performed for the $AVL(1 - strip \cdot T)$ elements that remain.

Fig. \ref{fig:strip} also illustrates three mainstream implementations of vector processing, highlighting their approaches to handling strip increments. In AVX programming, the strip size is fixed to the capacity of vector registers in the micro-architecture, such as \texttt{XMM}, \texttt{YMM}, or \texttt{ZMM}. Here, hardware does not participate in strip-mining control; instead, programmers calculate the strip size manually, often using the \texttt{sizeof} function. Both SVE and RVV adopt vector-length agnostic approaches, requiring hardware feedback to report the current micro-architectural attributes. SVE uses the \texttt{incd} assembly to increment a  counter based on the register size, which can vary by hardware implementation, typically ranging from 128 to 2048 bits. Combined with the \texttt{whilelt} and branch assemblies with predication, strip-mining loops can be implemented through compiler auto-vectorization, eliminating the need for detailed micro-architectural knowledge. Additionally, RVV enforces VL constraints via a load balancing unit controlled by the \texttt{vsetvli} instruction. Programmers specify a desired level of parallelism, and then the load balancing unit adjusts it to suit the current architecture. Finally, the adjusted strip size is used to update the strip counter.

Strip-mining arises as a trade-off among data parallelism, hardware utilization, and software feasibility. However, when applied to long vector processing, it encounters two primary challenges. Firstly, the strip size is constrained by the capacity of vector registers. Although RVV employs power-of-two register group multipliers to mitigate this limitation, the overhead of strip-mining -- such as increment/decrement, comparisons, and branching -- is not negligible in long vector scenarios. Furthermore, an improper strip size can lead to fragmented memory access patterns. Secondly, vector processing is limited by the number of available vector registers, especially when configuring a bloated LMUL. When multiple operations are bundled within a strip-mining loop, frequent register spilling and filling can occur. These two challenges not only increase memory traffic but also degrade overall throughput, further exacerbating inefficiencies in long vector processing. Compared to conventional vector programming, UVP leverages its support for non-power-of-two RG, effectively alleviating the issues of register under-utilization and excessive strip-mining instructions. Even when the VL is a power of two, UVP still demonstrates advantages due to reduced instruction fetch overhead and fewer pipeline stalls.

\begin{table*}[t]
  \centering
  \setlength{\tabcolsep}{8mm}{}
  \renewcommand\arraystretch{1.4}
  \caption{Instruction Encoding for UVP}
  \label{tab:inst_encode}
  \begin{threeparttable}[b]
  \begin{tabular}{rll}
    \toprule
    Bit field                & Name                & Description  \\ \hline
    \midrule
    {[}31:25{]}              & funct7{[}6:0{]}     & Operations under the category.                          \\ \hline
    {[}24:23{]}              & vew{[}1:0{]}        & Element width of the vector.                            \\ \hline
    {[}22{]}                 & vmask               & Whether to read predicate registers before computation. \\ \hline
    {[}63:52{]}, {[}21{]}    & vd\_head{[}12:0{]}  & Starting number of destination vector register.         \\ \hline
    {[}51:42{]}, {[}20:18{]} & vs2\_head{[}12:0{]} & Starting number of the second source vector register.   \\ \hline
    {[}41:32{]}, {[}17:15{]} & vs1\_head{[}12:0{]} & Starting number of the first source vector register or scalar register number.    \\ \hline
    {[}14:12{]}              & funct3{[}2:0{]}     & Categories of vector instructions.                      \\ \hline
    {[}11:7{]}               & rs\_avl{[}4:0{]}    & Scalar register number storing AVL.                     \\ \hline
    {[}6:0{]}                & opcode{[}6:0{]}     & Operation code of current instruction.                  \\ \hline
    \bottomrule
\end{tabular}
\end{threeparttable}
\vspace{-0.5cm}
\end{table*}

\section{Unlimited Vector Processing}
\label{UVP}
The formalization of the proposed ISE is divided into three subsections. First, we present the instruction encoding of the ISE, and introduce new instructions tailored to the unique demands of WBP. Next, we describe our new programming model and the organization of RG. Finally, we illustrate the practicality of our approach through two representative kernels in WBP, showcasing the effectiveness of the ISE and programming model.
\subsection{ISE Specifications}

Table \ref{tab:inst_encode} presents one possible instruction format of UVP. The UVP instruction bit width is 64 bits to accommodate the expanded physical vector register files (VRFs). While the instruction fetch may require additional clock cycles, this overhead can be offset by the SIMD nature. Details of the instruction formats are as follows.
\subsubsection{Operation Code}
The UVP ISE is designed based on the RISC-V ISA, which provides custom operation codes that allow developers to add new instructions while maintaining compatibility with the standard set. Our ISE utilizes \textit{custom-1} and \textit{custom-2} for this purpose.
\subsubsection{AVL Register}
A general-purpose scalar register is used to store the VL, with the register number encoded in the instruction. Unlike \texttt{vsetvli}, we simply reuse the \texttt{add} or \texttt{addi} instructions to inform hardware of the AVL for the entire vector. For example, prior to executing a \texttt{uvp\_add} instruction with \texttt{a0} as the AVL register, the compiler emits configuration assembly as shown below: 
\begin{equation}
\begin{aligned}
    \texttt{addi a0, a0, <immediates>} \\
    \texttt{uvp\_add v0, v0, v1, a0}.
\end{aligned}
\end{equation}
Here, \texttt{v*} represents vector registers. Furthermore, the compiler can also optimize the live interval of scalar registers when the VL remains constant or is modified by basic operations across instructions.
\subsubsection{Instruction Categories}
The instructions can be categorized based on the types of source and destination vectors. For source operands, one operand can be either a vector or a scalar. For destination operands, results can be written to either VRFs or dedicated predication registers. 
\subsubsection{Predication Registers}
For hardware efficiency, the ISE only supports one predication register. This design eliminates the need for complicated data widening and narrowing logic in the data pipeline, which can arise from bit-width differences between vector and mask elements. The predication register length matches the total number of bytes that all VRFs can process.
\subsubsection{Vector Element Width}
The ISE currently supports only \texttt{char} and \texttt{short} data types, as their bit widths are sufficient for quantization in WBP. Silicon previously allocated to less frequently used units, such as floating-point units and high bit-width multipliers, can be redirected to enhance data parallelism.
\subsubsection{Supported Operations}
In addition to basic fixed-point arithmetic, mask and reduction instructions, the proposed ISE allows different VLs between source and destination vectors through two new instructions: \texttt{uvp\_gather} and \texttt{uvp\_scatter}. Unlike \texttt{vrgather} in RVV, which operates under the same VL and can lead to low VRF utilization in certain LMUL configurations, these \textit{asymmetric} instructions maximize VRF efficiency when there is a substantial VL difference between source and destination registers. For instance, scattering often results in longer vectors:
\begin{equation}
\texttt{vd[vs2[i]] $\leftarrow$ vs1[i]}. 
\end{equation}
The \texttt{uvp\_scatter} instruction eliminates the need to pad vector registers associated with \texttt{vs1} and \texttt{vs2} to match the VL of \texttt{vd}. Conversely, gathering often results in shorter vectors: 
\begin{equation}
\texttt{vd[i] $\leftarrow$ vs1[vs2[i]]}. 
\end{equation}
In this case, registers related to \texttt{vd} and \texttt{vs2} are cropped to a shorter VL.
\subsubsection{Vector Registers}
For hardware implementation, single-port static RAMs (SRAMs) can be used instead of registers or register files due to their advantages: (i) SRAMs offer higher cell density, saving silicon area when the number of physical registers is large; and (ii) it is unlikely that more than two read/write requests will occur on an SRAM in a single clock cycle. In case of burst requests, implementing a request queue with back-pressure is more efficient. The depth of VRFs is up to $2^{13}$ as specified in Table \ref{tab:inst_encode}, but it can be further expanded by using \texttt{uvp\_vsetcsr} instruction to write extra bits into control and status registers (CSRs).

\subsection{Programming Model}

\begin{figure}[!t]
  \centering
  \includegraphics[width=\linewidth]{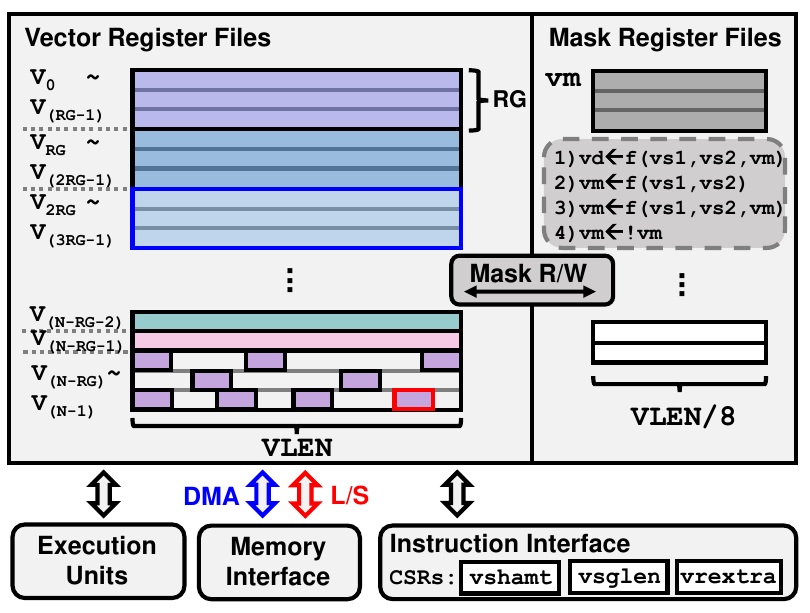}
  \vspace{-0.6cm}
  \caption{The programming model of our proposed unlimited vector processing (VLEN: Vector length of a single vector register).}
  \label{fig:progmodel}
  \vspace{-0.5cm}
\end{figure}

\begin{table}[t]
  \centering
  \caption{RG organization of asymmetric instructions}
  \label{tab:rg}
  \begin{threeparttable}[b]
  \begin{tabular}{r p{2.2cm} p{2.2cm} p{1cm} }
    \toprule
    & \texttt{vd} & \texttt{vs1} & \texttt{vs2}\\
    \midrule
    \texttt{uvp\_gather} & $V_{N-RG-2}$ & $V_{N-RG}\sim V_{N-1}$ & $V_{N-RG-1}$ \\
    \texttt{uvp\_scatter} & $V_{N-RG}\sim V_{N-1}$ &$V_{N-RG-2}$ & $V_{N-RG-1}$\\
    \bottomrule
\end{tabular}
\end{threeparttable}
\vspace{-0.5cm}
\end{table}

To address the issue outlined in Section \ref{Motivations}, we propose a programming model for UVP that is tailored for high-throughput vector processing. UVP is achieved by addressing two key constraints: the granularity of vector register grouping and the hardware strip-mining. Unlike traditional ISAs, UVP offers greater flexibility in vector register grouping, allowing for non-power-of-two groupings. This enables programmers to access arbitrary consecutive vector registers within a single instruction, thereby reducing the execution overhead of strip-mining loops. Additionally, UVP shifts strip-mining from software to hardware by leveraging the AVL register in instruction encoding, which reduces the frequency of memory accesses through finer-grained register management. However, UVP introduces challenges for compiler implementation. While mainstream ISAs exhaust registers and sub-registers with relatively low complexity, UVP's flexibility necessitates more sophisticated handling. Although we have implemented several passes in an LLVM compiler to address these challenges, the details lie beyond the scope of this study and will be explored in future work.

An overview of the proposed programming model is illustrated in Fig. \ref{fig:progmodel}. The key features of the programming model are detailed as follows.

\subsubsection{Vector Register Files}
The VRF usage of UVP offers greater flexibility, with RGs dynamically adjusted based on different instructions. RG is determined offline by two factors: the starting register number, $RG_{head}$, allocated by RA algorithms of the compiler, and the vector data type, defined as $RG_{type}=L \cdot VEW$, where $L$ is the VL specified by the programmer through high-level language. At compile time, RGs are allocated by calculating its range from $RG_{head}$ to $RG_{tail} = RG_{head} + RG_{type}/VLEN$. During runtime, the hardware accesses vector registers based on the AVL register specified in the instruction encoding, which allows for cases where the user-defined size exceeds the actual runtime vector size. This design introduces additional complexity in hardware implementation, as UVP requires managing two degrees of freedom ($RG_{head}$ and $RG_{type}$), in contrast to traditional ISAs, which typically rely on only one ($RG_{head}$).

RG organization varies depending on the instruction type. Symmetric instructions execute element-wise computations, where the destination registers have the same length as the source registers. Operands in symmetric instructions are densely arranged, making them well-suited for direct memory access (DMA). This dense arrangement allows efficient vector load/store operations, typically performed before or after strip-mining. As illustrated in Fig.~\ref{fig:progmodel}, the source and destination RG in symmetric instructions occupy the same number of vector registers. In contrast, asymmetric instructions involve differing lengths between source and destination registers. The source operands to be gathered or scattered are sparsely arranged. In such scenarios, scalar load/store instructions from the scalar core are more efficient than vector load/store operations. Table~\ref{tab:rg} presents the possible RG organizations for asymmetric instructions, aligned with Fig.~\ref{fig:progmodel}.

\subsubsection{Mask Register Files}
Mask register files (MRFs) store predication results, with a size one-eighth that of the VRF, holding mask content in bits. For implementation simplicity, there is a single mask register, \texttt{vm}, which holds the latest predication result. MRF contents are inaccessible via the memory interface and can only be modified by specific instructions. For example, a compare instruction (e.g. \texttt{uvp\_seq}) can operate with or without input from mask registers, and its result can be written back to either the VRF or MRF. Additionally, the mask register can reverse predications using the \texttt{vmnot} instruction, enabling efficient handling of vector \texttt{if-else} conditions.

\subsubsection{Control and Status Registers}
Control and status registers (CSR) provide additional information for UVP through a dedicated instruction interface. For instance, \texttt{vsglen} specifies the vector length of destination registers in scatter and gather instructions, addressing the asymmetric nature of these instructions. The \texttt{vrextra} register extends instruction encoding space to accommodate the indices of vector registers when the current instruction format cannot specify a sufficient number of vector registers. For computation, \texttt{vshamt} denotes the shift amount of operands to be applied to operands before or after the computation, a feature which will be further clarified in Section \ref{lane}. Although CSR configuration instructions might be required prior to each vector instruction, the overhead is minimal and easily amortized due to the efficiency of long vector processing.

In summary, we define the syntax of the proposed programming model in the C programming language as follows:
\begin{equation}
[V_{dest}=]f_{uvp1}(V_{src} | Addr, [B_r,] AVL), 
\end{equation}
\begin{equation}
[V_{dest}=]f_{uvp2}(V_{src1}, V_{src2}, [V_{src3}, B_r, B_w,] AVL), 
\end{equation}
\begin{equation}
f_{uvp3}(V_{dest1}, [V_{dest2},] V_{src1}, V_{src2}, [V_{src3}, V_{src4}, B_r,] AVL).
\end{equation}
Here, the three types of $f_{uvp}$ correspond to load/store operations, normal arithmetic instructions, and complex/asymmetric instructions, respectively. The optional boolean flags $B_r$ and $B_w$ indicate whether the hardware should read from or write to MRFs. 

  

\begin{figure*}[!t]
  \centering
    \includegraphics[width=0.98\linewidth]{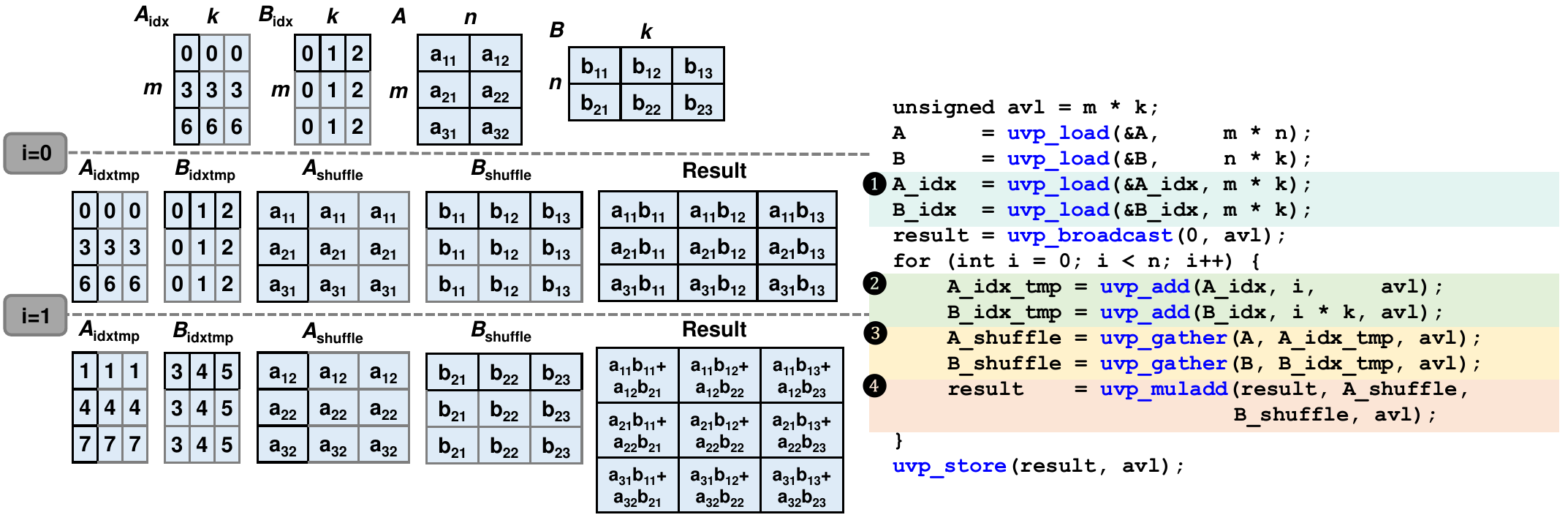}
  \vspace{-0.3cm}
  
  \caption{A \textit{matmul} kernel multiplying a $3\times2$ matrix by a $2\times3$ matrix, requiring two iterations.}
  \label{fig:matmul}
  \vspace{-0.5cm}
\end{figure*}

\subsection{Kernel Examples}
\label{section:kernel}
To showcase the advantages of UVP, we present two kernel examples commonly used in WBP. Leveraging large vector registers and flexible vector organizations, these kernels are rewritten to achieve greater efficiency and improved performance.

\subsubsection{Matrix Multiplication}

Let $\boldsymbol{a}_i$ denote the $i$-th column vector of the matrix $\boldsymbol{A}_{m\times n}$, and $\boldsymbol{b}_j$ denote the $j$-th row vector of the matrix $\boldsymbol{B}_{n\times k}$. The resulting matrix $\boldsymbol{C}_{m\times k}$ from their multiplication can be expressed as
\begin{equation}
    \label{eq:matmul}
    \boldsymbol{C} = \sum_{i} \left( 
        \overbrace{\begin{bmatrix} \boldsymbol{a}_i & \boldsymbol{a}_i  & \cdots & \boldsymbol{a}_i \end{bmatrix}}^{n}
        \odot 
        \left. \begin{bmatrix} \boldsymbol{b}_i \\ \boldsymbol{b}_i \\ \vdots \\ \boldsymbol{b}_i \end{bmatrix} \right\} \,_{n}
        \right),
\end{equation}
where $\odot$ is the Hadamard product. 

As shown in Fig.~\ref{fig:matmul}, matrix multiplication (\textit{matmul}) on UVP can be performed in the following steps. \textbf{\ding{182} Permutation matrix generation.} Two permutation matrices can be generated as follows:
\begin{equation}
    \boldsymbol{\Pi_{A}} = \overbrace{\begin{bmatrix} \boldsymbol{\pi_{A}} & \boldsymbol{\pi_{A}}  & \cdots & \boldsymbol{\pi_{A}} \end{bmatrix}}^{n},
\end{equation}
\begin{equation}
    \boldsymbol{\Pi_{B}} = {\overbrace{\begin{bmatrix} \boldsymbol{\pi_{B}} & \boldsymbol{\pi_{B}}  & \cdots & \boldsymbol{\pi_{B}} \end{bmatrix}}^{n}}^{T},
\end{equation}
where $\boldsymbol{\pi_{A}}$ and $\boldsymbol{\pi_{B}}$ represent the column vector and row vector of matrices $\boldsymbol{\Pi_{A}}$ and $\boldsymbol{\Pi_{B}}$, respectively:
\begin{equation}
    \boldsymbol{\pi_{A}} = {\begin{bmatrix} 0 & n & \cdots & (m-1)n \end{bmatrix}}^{T},
\end{equation}
\begin{equation}
    \boldsymbol{\pi_{B}} = \begin{bmatrix} 0 & 1 & \cdots & k-1 \end{bmatrix}.
\end{equation}
The permutation matrices can either be stored in advance or generated dynamically by a sequence of fundamental instructions, given their simplicity. \textbf{\ding{183} Permutation matrix update.} Within the calculation loop, the permutation matrices are updated by adding scalar values corresponding to the iteration variables. \textbf{\ding{184} Element shuffling.} Gather instructions are utilized alongside the updated permutation matrices to extract specific elements from the original data matrices. These extracted elements are reorganized to form new matrices aligned for the subsequent computation. \textbf{\ding{185} Resulting matrix accumulation.} At the end of each computation loop, the resulting matrix is updated by accumulating the products of the shuffled matrices. Notably, the permutation matrix and the shuffled matrix in steps \ding{184} and \ding{185} can be reduced into a vector due to their identical row-wise or column-wise structure. This further reduces the kernel runtime.

In this case, UVP demonstrates a more flexible approach to data computation. The size of the RG is directly influenced by the programmer-specified vector length (\texttt{avl}), provided the total number of source and destination vector registers remains within the overall VRF capacity. Compared to traditional \textit{matmul} \cite{armmatmul, git-ara}, which has time and space complexities of $\mathcal{O}(N^2)$ and $\mathcal{O}(N)$, respectively, UVP achieves higher throughput by utilizing large VRFs. This optimization effectively reverses the time and space complexity trade-off, significantly enhancing performance in scenarios that require intensive vector operations.

\begin{figure*}[!t]
  \centering
    \includegraphics[width=0.88\linewidth]{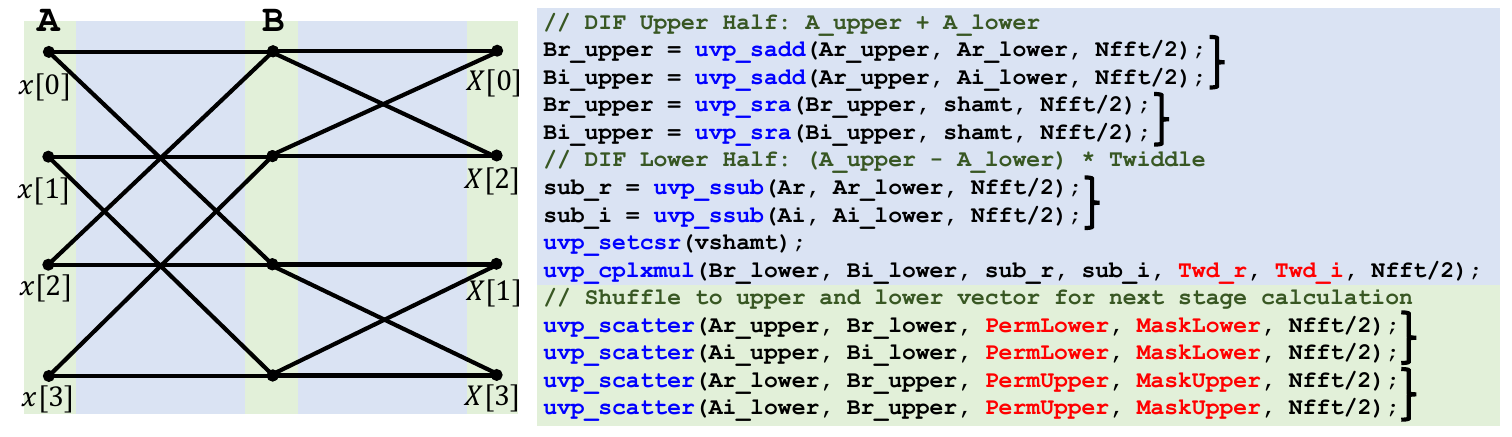}
  \vspace{-0.3cm}
  \caption{A DIF \textit{fft} kernel featuring saturation and complex multiplication instructions within each stage of the dataflow. Variables in red denote prepared permutation vectors for gather and scatter operations.}
  \label{fig:fft}
  \vspace{-0.5cm}
\end{figure*}

\subsubsection{Fast Fourier Transform}
Existing kernel implementations of fast Fourier transform (\textit{fft}) \cite{git-ara} have already optimized the computation process by calculating stage by stage, employing a ``shuffle-butterfly-shuffle'' strategy within each stage instead of recursively calling a half-point \textit{fft}. However, these 
implementations face several limitations:
\begin{itemize}
    \item \textbf{Vector Register Size Constraints.} The limited size of existing vector registers restricts the maximum FFT point size. In wireless communication, where FFT size are often large ($N_{fft}\geq2048$), this becomes a significant bottleneck.
    \item \textbf{Complex Number Representation.} Signal processing relies heavily on data represented as complex numbers, which necessitates efficient complex multiplication. The current implementations lack the ability to reduce instruction counts and efficiently chain the dataflow for such operations.
    \item \textbf{Data Saturation Needs.} Signal processing frequently requires data to saturate to the maximum and minimum values representable in small data types to minimize system error. This demands tailored operations for saturation arithmetic apart from addition and subtraction.
\end{itemize}

\begin{figure*}[!t]
  \centering
  \includegraphics[width=\linewidth]{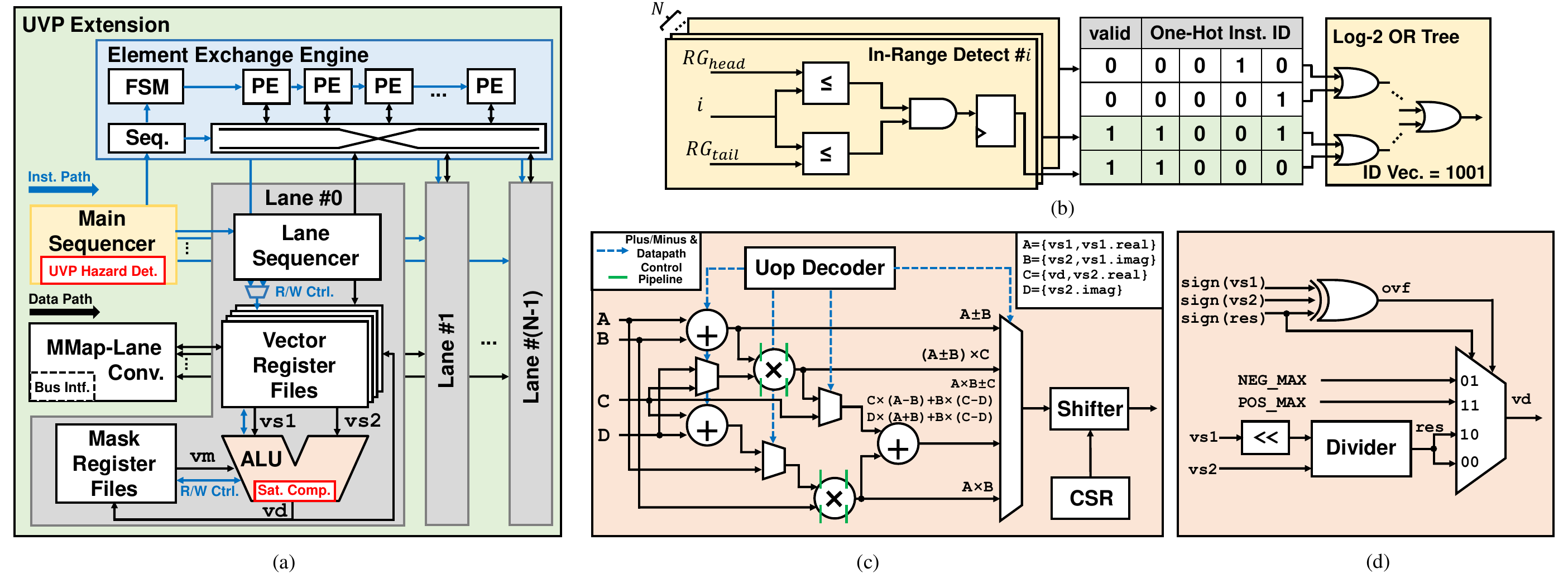}
  \vspace{-0.7cm}
  \caption{The overall top view of the UVP microarchitecture and key subdesigns. (a) The overall top view of the UVP microarchitecture, depicting the arrangement and interaction of core modules. (b) Hazard detection scheme of within the main sequencer, focusing on handling vector operand dependencies. (c) Complex arithmetic unit within the ALU, equipped with a final-stage shifter for precise operations. (d) Peripheral logic design surrounding a divider, dedicated to overflow detection and calculation.}
  \label{fig:arch}
  \vspace{-0.5cm}
\end{figure*}

Fig.~\ref{fig:fft} illustrates a decimation-in-frequency (DIF) \textit{fft} kernel with several enhancements. The expanded VRFs allow for the computation of larger-point FFTs. Scatter instructions, combined with permutation vectors and mask flags (marked in red), are employed to re-organize vectors after the butterfly calculation for the next stage of computation. These vectors can either be pre-stored or calculated at runtime, depending on throughput requirements. The instruction count is further optimized by combining the real and imaginary parts of the signal into one single vector. During the calculation, a specialized complex multiplication instruction, \texttt{uvp\_cplxmul}, reduces the instruction count from six (four multiplications and two add/subtract instructions) to a single instruction, streamlining the computation and enhancing dataflow efficiency. Moreover, the \texttt{vshamt} register can be pre-configured using the \texttt{uvp\_setcsr} instruction to handle data overflow during calculations, which eliminates the need for additional overflow management operations.

\section{Overall Microarchitecture}
\label{section:Hardware}
Building upon the proposed programming model, we design an architecture that addresses both hardware strip-mining and WBP requirements. As shown in Fig. \ref{fig:arch}(a), the microarchitecture is divided into four components. (i) The main sequencer is responsible for routing instructions to the appropriate execution units and stalling the pipeline when there are RG conflicts. (ii) The memory-map-to-lane conversion module contains address generation logic, converting between VRF SRAM addresses and the memory-mapped address organization. A configurable bus interface, compliant with the Advanced eXtensible Interface (AXI), allows the design to access memories actively or passively. (iii) Scalable lanes are dedicated to element-wise computations with the ALUs accessing in-situ VRFs and MRFs. (iv) The element exchange engine (EXE) enables efficient permutations of vector elements across the lanes.

\subsection{Control Path for UVP}
The control path of the UVP begins at the main sequencer, traverses the execution units within lanes and the EXE, and concludes at VRFs, as depicted by the blue lines in Fig. \ref{fig:arch}(a). Unlike traditional strip-mining techniques, where the instruction decoder determines the maximum number of vector elements to process per instruction based on the size of physical registers, vector element type, and the LMUL (if applicable) \cite{cavalcante2019ara}, UVP achieves hardware strip-mining by pipelining the AVL, decoded from the general-purpose register index in the RISC-V ISA, within the main sequencer. Afterwards, the AVL is further divided within the lane sequencer, with the VL for each lane $i$ calculated directly using the pipelined AVL register:
\begin{equation}
VL_{i} = \lfloor AVL/N_{lane} \rfloor + N_{tail,i}, 
\end{equation}
where $N_{lane}$ denotes the number of lanes in the UVP extension, and $N_{tail,i}$ is defined as:
\begin{equation}
N_{tail,i} = 
\begin{cases}
0 & (AVL \mod N_{lane}) \geq i  \\
1 & (AVL \mod N_{lane}) < i
\end{cases}.
\end{equation}
The calculated result determines the number of micro-operations (uops) each execution unit must perform. This approach eliminates the VL constraint specified in the RVV specification, enabling UVP to directly execute vector processing based on the programmer-defined AVL.

The control signals proceed to fetch operands for both the lanes and the EXE for $VL_{i}$ iterations. Within each lane, the lane sequencer requests operands from  the VRFs and MRFs. The request logic utilizes a multiple-input-single-output arbiter, which handles requests from the in-situ lane as well as the processing elements (PEs) in the EXE. Along with the fetched data, the control signals are further broken down into smaller uops for multi-pipeline execution units. Upon execution, the control signals direct the results back to the VRFs or MRFs. Similarly, for EXE, a finite state machine (FSM) is triggered by the internal sequencer, which itself is activated by the main sequencer. The PEs within the EXE request and write back operands through the same arbiter, ensuring efficient operand management across lanes. 

Control signals also facilitate the coordination of lanes and the EXE to execute reduction operations, which are carried out in two stages. In the first stage, intra-lane reduction is performed, where each lane reduces its $VL_{i}$ elements as part of normal symmetric operations. In the second stage, the EXE handles inter-lane reduction over $\log_{2}{N_{lane}}$ iterations. During iteration $n$ (where $n=0,1,...,\log_{2}{N_{lane}}-1$), the EXE fetches operands from lane $(m2^{n+1}+2^{n})$ and writes back to lane $m2^{n+1}, m=0,1,...,N_{lane}/2^{n+1}$. After completing these steps, the final reduction result is written back to the VRFs in lane 0.

\subsection{Hazard Detection Logic for UVP}

Another challenge, apart from hardware strip-mining, is hazard detection. Hazards arise when operands are repeatedly accessed in consecutive instructions, requiring hardware to ensure that the subsequent instructions delay any read or write operations until the previous instruction finishes its write-back. Given a total of $N$ physical vector registers, the sum of available logic registers across different LMULs in RVV is calculated as $\sum_{n_{lmul}=0}^{3}N/2^{n_{lmul}}$, since only registers $V_{k2^{n_{lmul}}}, 0 \leq k \leq N/2^{n_{lmul}}-1$ can serve as the head vector of an RG. However, in UVP, RGs can be arbitrarily assigned, meaning any vectors can act as the head or tail of an RG. This flexibility results in a total of $2^{N-1}$ possible logic register configurations. 

In UVP, hazard detection complexity surpasses that of power-of-two RG configuration. As shown in Fig. \ref{fig:arch}(b), each physical register is examined to verify whether it falls within the specified range, defined by the head and tail of an RG. Validity information is stored in a table along with a one-hot encoded instruction identity (Inst. ID) assigned by an instruction monitor, which manages up to $N_{ID} = 8$ concurrent instructions. This table feeds into a $\log_{2}{N}$-level OR tree, using a validity mask to condense information into an ID vector that flags instruction occupation. The ID vector then updates an $N_{ID} \times N_{ID}$ hazard table, which informs subsequent modules of potential pipeline stalls. Validity and IDs clear automatically when the instruction completes. This approach requires $2N$ more comparators than the traditional data hazard recording methods \cite{cavalcante2019ara}, but it enables a more adaptable RG strategy.

\subsection{Computation within Lanes}
\label{lane}
A lane operates as a subordinate packed-SIMD core, consisting of a lane sequencer that issues uops, temporary storage for operands within mask and vector register files, and an ALU that executes parallel computations. Execution units are further enhanced to support complex multiplication and saturation division, specifically tailored for WBP. As depicted in Fig. \ref{fig:arch}(c), a complex arithmetic unit (CAU) includes multiple datapaths designed to handle multiplication-related operations. A uop decoder interprets the instructions and generates control signals for adders and datapath multiplexers. Supported instructions include addition/subtraction, multiplication followed by addition/subtraction,  addition/subtraction followed by multiplication, and dedicated complex multiplication. To improve timing performance, two pipeline registers are placed before and after the multipliers. Calculations are performed in full precision by default. A configurable truncate module, managed via \texttt{vshamt} CSR, allows operations in fixed-point formats to meet precision and application-specific requirements.

Fix-point division requires careful consideration of two issues: (i) Direct division can lead to underflow, resulting in imprecise number representation and loss of fractional details; and (ii) the quotient may overflow when the dividend is sufficiently small. For a number represented with an $m$-bit integer part and an $n$-bit fractional part, noted as $m$Q$n$, the full-precision calculation of between a divisor of $a$Q$b$ and a dividend of $c$Q$d$ produces a quotient format of $(a+d+1)$Q$(b+c)$. To mitigate underflow, an additional left shift of $s$ can be applied to the divisor using \texttt{vshamt} CSR. This shift increases the fractional precision of the quotient at the cost of reducing its integer precision, yielding a quotient format of $(a+d+1-s)$Q$(b+c+s)$. As illustrated in \ref{fig:arch}(d), a configurable left shift logic is implemented before the divisor input. To handle overflow, a detection mechanism is implemented using an XOR gate. The gate takes the sign bits of the divisor, dividend, and unsaturated quotient as inputs. Based on the sign of the unsaturated quotient, the output is saturated to its minimum or maximum value accordingly.

\subsection{Element Exchange Engine}

\begin{figure}[!t]
  \centering
  \subfloat[]{
    \includegraphics[width=\linewidth]{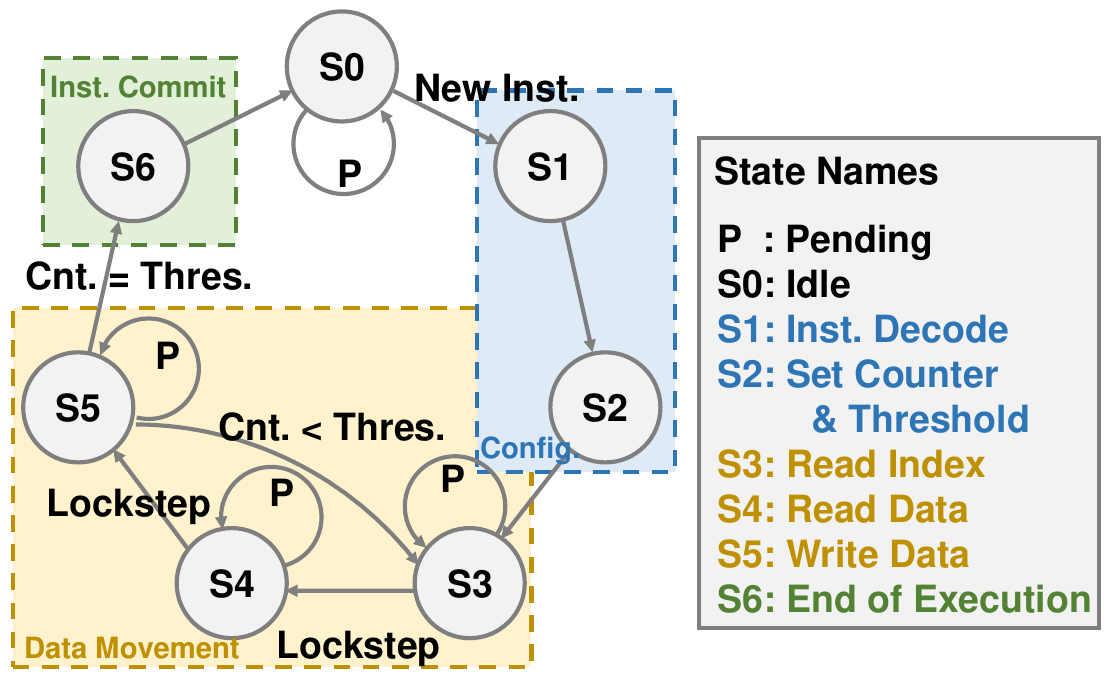}
    \label{fig:shuffleFSM}
  }
  \vspace{-0.4cm}
  \subfloat[]{
    \includegraphics[width=0.98\linewidth]{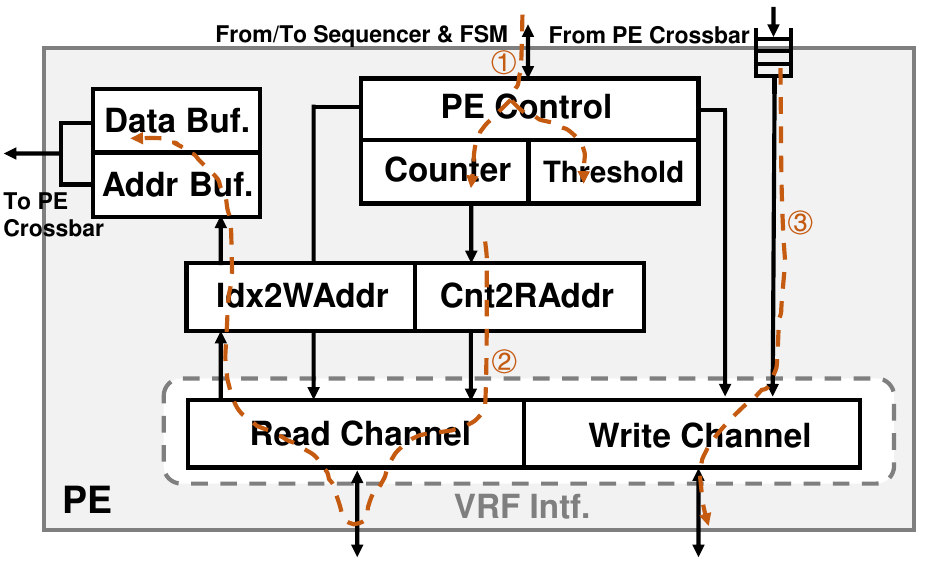}
    \label{fig:shufflePE}
  }

  \caption{Details inside EXE. (a) A finite state machine that controls the EXE. (b) Hardware design of shuffle processing elements.}
  \label{fig:shuffle}
  \vspace{-0.6cm}
\end{figure}

The EXE in the UVP extension enables element gather and scatter operations without requiring memory access. While some vector processing microarchitectures support only limited shuffling of vector elements -- often at the granularity of a physical register to keep hardware design simpler -- this limitation can lead to frequent memory access as the VL increases. In such cases, vectors must first be stored to memory and then reloaded using memory-based gather/scatter instructions. In contrast, the EXE can gather and scatter elements to any positions within VRFs, facilitating efficient data manipulation across elements without engaging memory.

Highlighted in Fig. \ref{fig:arch}(a), the EXE consists of an EXE sequencer, an FSM, PEs for data shuffling, and an interconnect. Fig. \ref{fig:shuffle} further details the processing sequence, where the FSM oversees the EXE's operations, prioritizing data integrity over throughput to mitigate hazards compared to a pipelined approach. The FSM, depicted in Fig. \ref{fig:shuffle}\subref{fig:shuffleFSM}, comprises three main stages: configuration, data movement, and instruction commit. In the configuration stage, the FSM decodes instructions and determines the number of data movements per PE as the EXE sequencer receives a new command. Counter and threshold values are reset and loaded into the PEs. During data movement, each PE loads indices and elements for shuffling and transmits them to the interconnect module, which then guides the data to the appropriate PEs. While receiving and writing back the shuffled data into the VRF, the FSM ensures all the PEs operate in lockstep before advancing. Finally, in the instruction commit stage, a complete signal is sent to the main sequencer, which updates the instruction monitor and data hazard table.

Based on the FSM design, each PE is implemented to fetch and write back vector elements in a cyclic \ding{172}$\rightarrow{}$\ding{173}$\rightarrow{}$\ding{174}$\rightarrow{}$\ding{172} procedure,  as illustrated in Fig. \ref{fig:shuffle}\subref{fig:shufflePE}. A PE control with a counter and a threshold register is included to manage all subsequent modules according to the FSM specifications. The write and read channel interface with vector operands, with operand requests managed by an arbiter within each lane. Additionally, a counter-to-read-address (\textit{Cnt2RAddr}) module is included to sequentially read elements from both the index and target vector, beginning from the first element of an RG. Addressing differs between VRF SRAMs and the index vector contents: VRF SRAMs are accessed within the lane, while index vectors pertain to the RG and span multiple lanes. Therefore, two conversion modules, \textit{Idx2PE} and \textit{Idx2WAddr}, are implemented to direct data to the correct PE and to compute the SRAM address within the corresponding lane, respectively.

\section{Experiment Results}
\label{section:Experiment}
In this section, we evaluate our extension at both the software and hardware levels. At the software level, we compare kernels and instruction counts introduced in Section \ref{section:kernel} with the open-source RVV processor Ara \cite{cavalcante2019ara} under various configurations. At the hardware level, we compare our proposed architecture with other state-of-the-art vector processors, providing a detailed architectural breakdown to highlight specific design benefits.

\subsection{Kernel Comparison}
\begin{figure}[!t]
  \centering
  \includegraphics[width=\linewidth]{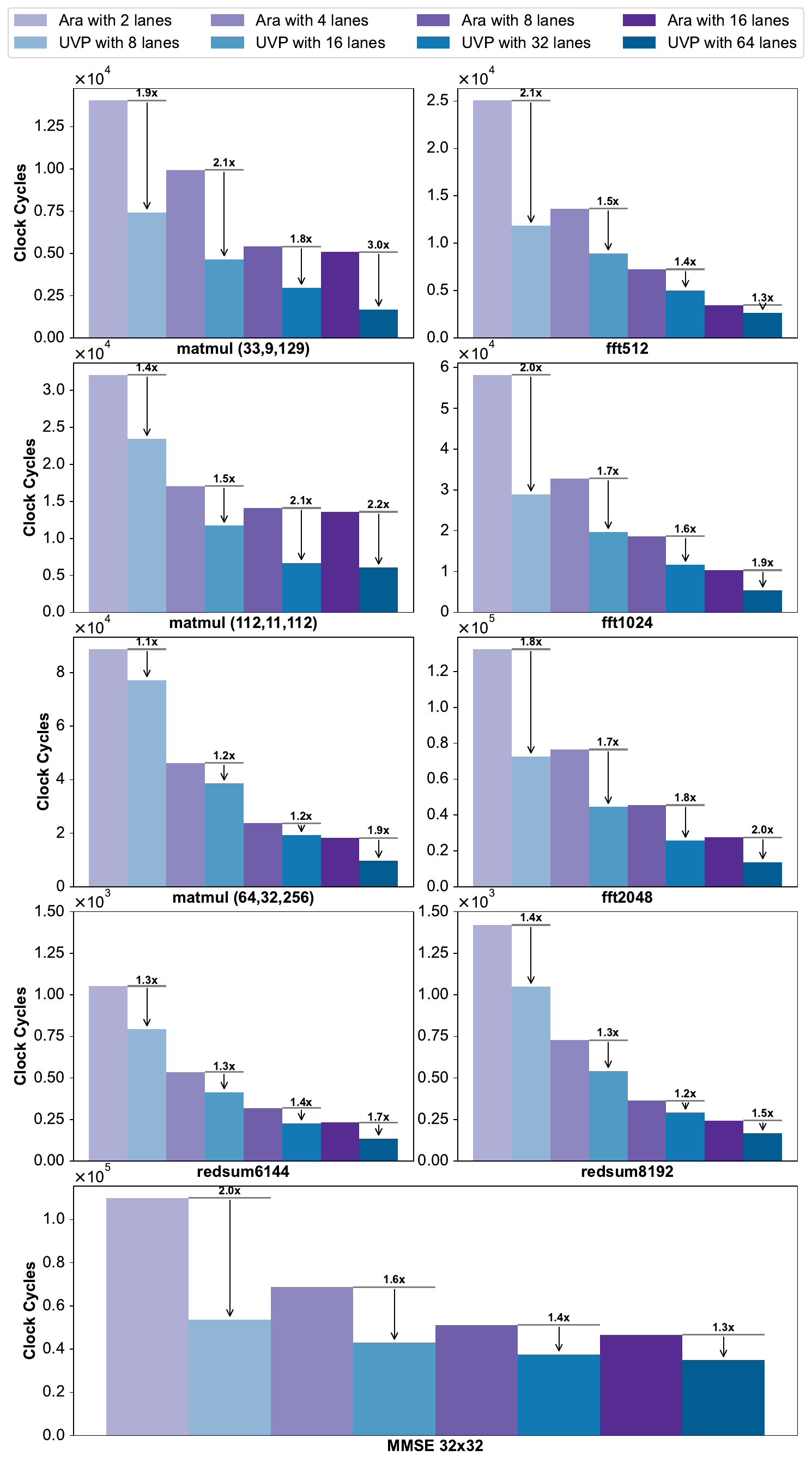}
  \vspace{-0.7cm}
  \caption{Comparison of UVP with Ara in the number of clock cycles across various configurations and kernels. Note that we compare the $L$-lane Ara with the $4L$-lane UVP (as shown in the legend, with pairs organized in columns) to ensure a fair comparison, given the equal number of execution units. }
  \label{fig:kernel}
  \vspace{-0.5cm}
\end{figure}

We select \textit{imatmul}, \textit{fft}, and \textit{redsum} kernels from \cite{git-ara} as benchmarks to compare with our design. In addition, a minimum mean square error (MMSE) kernel was implemented to compare with \cite{acevedo20245g} in multiple-input, multiple-output (MIMO) systems. To align with the execution unit configurations, we modified the element type of the vectors to \texttt{int16} in each kernel. At the hardware level, the number of lanes in UVP is set to be four times that of Ara, due to the wider SRAMs and the increased DLP of Ara. Clock cycles are measured using Synopsys VCS and QuestaSim, respectively. Fig. \ref{fig:kernel} shows the kernel speedup across four lane configurations. \textit{matmul (m,n,k)} represents an $\boldsymbol{A}^{m \times n} \times \boldsymbol{B}^{n \times k}$ matrix multiplication, \textit{fftN} represents a $N$-point FFT, and \textit{redsumM} indicates a reduction summation over $M$ elements. For MMSE estimation, we adopt a $32 \times 32$ window size, consistent with \cite{acevedo20245g}.

\subsubsection{Matrix Multiplication}
The speedup for \textit{matmul} ranges from 1.1$\times$ to 3.0$\times$, with the greatest improvements observed when the matrix dimensions are not powers of two, achieving at least 1.4$\times$ speedup. In the proposed \textit{matmul} kernel example, RGs under different AVL ($m \times k$) configurations are more compact compared to the power-of-two RGs. This results in fewer strip-mining loops and switches, as well as register vacancies. An additional source of speedup lies in the loop iteration strategies: Ara’s kernel operates at a complexity of $\mathcal{O}(m \times k / b^2)$, while the UVP kernel achieves $\mathcal{O}(n)$ by trading off increased space complexity to store matrix elements. Here, $b$ denotes the block size used to configure \texttt{vsetvli} instructions.

\subsubsection{Fast Fourier Transform}
For the \textit{fft} kernel, flexible RGs and larger VRFs enable UVP to achieve a performance improvement ranging from 1.3$\times$ to 2.1$\times$. The performance gap becomes more significant when the input data for the butterfly computation in a stage exceeds the capacity of vector registers in Ara. In such cases, a potential solution would be to call the most suitable kernel for the current hardware configuration and recursively calculate the FFT for larger points. However, this divide-and-conquer approach would introduce additional memory accesses. Under UVP, more data can be loaded into registers and grouped beyond the RVV specification, mitigating the need for excessive memory accesses and strip-mining loops. Additionally, the inclusion of complex multiplication instructions further reduces the clock cycles, contributing to overall improved performance.

\subsubsection{Reduction Summation}
Reduction operations are widely used in applications such as low-density parity-check (LDPC) decoding and signal correlation, and continue to be the subject of performance optimization efforts \cite{ScanGPU, lai2022efficient}. The fourth row of Fig.~\ref{fig:kernel} shows the performance of reduction summation for both power-of-two and non-power-of-two input lengths. Despite having four times the number of lanes compared to Ara -- requiring two additional stages of element permutation -- UVP consistently achieves at least a 1.2$\times$ speedup. This improvement is attributed to less strip-mining and configuration instructions, as well as the elimination of scalar reductions across vector instructions.

\subsubsection{MMSE Estimation}
The complexity of matrix inversion has prompted extensive optimization efforts across platforms \cite{chang2019case, bertuletti2023efficient}, with Cholesky decomposition being the most commonly adopted method. The bottom row of Fig.~\ref{fig:kernel} compares the performance of our implementation with that of \cite{acevedo20245g}, using a $32\times32$ matrix window for a $2\times2$ MIMO system. To avoid under-utilization of the VRF, multiple MMSE kernels are executed in parallel under larger lane configurations. UVP still outperforms the Ara implementation, owing to its flexible element reorganization for matrix operations and enhanced support for complex multiplications.

\begin{figure}[!t]
  \centering
  \includegraphics[width=\linewidth]{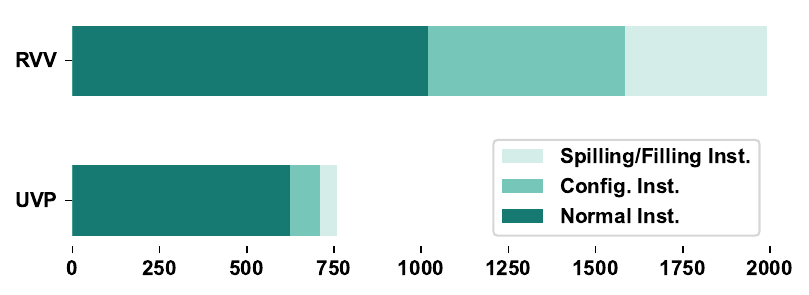}
  \vspace{-0.5cm}
  \caption{Fraction of instructions over the dynamic instruction count for the 2048-point FFT kernel.}
  \vspace{-0.5cm}
  \label{fig:inst_bkd}
\end{figure}

\subsection{Instruction Breakdown}
To further assess the impact of UVP, we analyze the dynamic instruction count of both RVV and UVP implementations. SystemVerilog Assertions (SVA) are integrated at the RTL level to track register spilling/filling and strip-mining-related instructions. As shown in Fig.~\ref{fig:inst_bkd}, UVP reduces the dynamic instruction count for the 2048-point FFT kernel to 38\% of that observed in RVV, underscoring three notable improvements in the proposed design. First, UVP reduces arithmetic instruction count by 1.6×, attributed to the introduction of dedicated complex multiplication instructions, which streamline computation. Second, configuration instructions — such as \texttt{vsetvl} in RVV and \texttt{uvp\_vsetcsr} in UVP — are significantly reduced due to UVP's flexible register grouping mechanism, which accommodates longer vectors and minimizes status switching overhead. Finally, UVP considerably decreases register spilling and filling instructions, owing to its asymmetric gather/scatter instructions and the kernel design that departs from the conventional divide-and-conquer approach, further enhancing computational efficiency.

\begin{table*}[t]
    \centering
    \caption{Implementation Results of the Proposed Vector Extension Compared with State-of-the-Art Vector Processors}
    \label{tab:impl_result}
    \begin{threeparttable}[b]
    \begin{tabular}{rccccccc}
    \toprule
                                & This work            & TC'24$^\dag$             & ICCAD'22$\ddag$            & TCASII'23$^\dag$              & TCASI'23                & TVLSI'23            & TVLSI'24              \\
                                & Lane$_{16}$Reg$_{32}$             & \cite{perotti2024ara2}   & \cite{cavalcante2022spatz} & \cite{perotti2023yun} & \cite{ottavi2023dustin} & \cite{zhao2023hipu} & \cite{wang2024speed}  \\\midrule\midrule
        \textbf{ISA Base \& Extensions} & Xuvp                 & GCV1.0                   & Zve32x                     & GCV0.9                & IMFAC$\times$16         & IMA+Vec/Mtx         & V1.0+Tensor           \\\midrule
        \textbf{Technology} (nm)& 40                   & 22                       & 22                         & 65                    & 65                      &  28                 & 28                    \\\midrule
        \textbf{Result Source}  & Synthesis            & Layout                   & Layout                     & Silicon               & Silicon                 & Silicon             & Synthesis             \\\midrule
        \textbf{Frequency} (MHz)& 400                  & 1350                      & 594                        & 280                   & 205                     & 1000                & 1050                  \\\midrule
        \textbf{Supply} (V)     & 1.1                  & 0.8                      & 0.8                        & 1.2                   & 1.2                     & 1.125               & 0.9                   \\\midrule
        \textbf{Core Area} (mm$^2$)&0.94                 & 0.95                     & 20.1                       & 6                     & 10                      & 72                  & 1.2 \\\midrule
        \textbf{Int. Formats} (bit) & 8, 16            & 8, 16, 32, 64            & 8, 16, 32                  & 8, 16, 32, 64         & 2, 4, 8, 16, 32         & 8                   & 4, 8, 16, 32, 64      \\
        \textbf{Best Int. Perf.} (GOPS@\textsc{Int8}) & 19.9       & 83.5                     & 285                        & 22.9                  & 58@\textsc{Int2}                 & 13000               & 737.9@\textsc{Int4}\\
        \textbf{Best Int. Area Eff.} (GOPS/mm$^2$) & 21.2 & 87.9                     & 14.2                       & 3.8                   & 5.8@\textsc{Int2}                     & 180.6               & 614.6@\textsc{Int4} \\
        \textbf{Best Int. Energy Eff.} (GOPS/W) & 250.5    & 376.0                    & 266                        & 100.5                 & 1152@\textsc{Int2}                    & 630                 & 1383.4@\textsc{Int4} \\\midrule
        \multicolumn{8}{c}{Normalized to 40nm technology\tnote{$\ast$}} \\\midrule
        \textbf{Norm. Area} (mm$^2$) & 0.94               & 3.1                      & 66.4                       & 2.3                   & 3.8                     & 146.9               & 2.4\\
        \textbf{Norm. Int. Perf.} (GOPS@\textsc{Int8}) & 19.9       & 45.9                     & 156.8                      & 37.2                  & 94.3@\textsc{Int2}               & 9100                & 516.5@\textsc{Int4} \\
        \textbf{Norm. Int. Area Eff.} (GOPS/mm$^2$) &21.2& 14.8                     & 2.4                        & 16.2                  & 24.8@\textsc{Int2}                    & 61.9                & 215.2@\textsc{Int4}\\\bottomrule
    \end{tabular}
    \begin{tablenotes}
        \footnotesize
        \item[$\dag$] A 4-lane configuration. 
        \item[$\ddag$] A Mempool$_{64}$Spatz$_4$ configuration.
        \item[$\ast$] All results are normalized to 40nm based on: frequency $\propto s$, area $\propto s^{-2}$, where $s$ is the scaling factor.        
    \end{tablenotes}
    \end{threeparttable}
    \vspace{-0.8cm}
\end{table*}

\subsection{Implementation Result}
We implement our proposed architecture at the RTL level and synthesize the design using Synopsys Design Compiler on the SMIC 40nm technology. 
Throughput performance is evaluated by issuing randomly generated instructions (with varied RGs and AVLs) to the extension using Synopsys VCS. Timing and power analyses are carried out under typical process, voltage, and temperature conditions (TT, 1.1V, 25\textcelsius). We label each configuration of our work as Lane$_l$Reg$_r$, where $l$ denotes the number of lanes and $r$ represents the depth of SRAM.

\subsubsection{Comparison with the State-of-the-Arts}
\label{big_table}
Table \ref{tab:impl_result} presents a comprehensive performance comparison with the state-of-the-art vector processors, normalizing performance and efficiency across different technologies for consistency. The proposed architecture is compared with the synthesis results of other processors to ensure a fair evaluation of computational capability.
In Ara2 \cite{perotti2024ara2} and Yun \cite{perotti2023yun}, lane-based vector processors supporting ratified RVV extensions with high-performance scalar cores are implemented and fabricated. Both Spatz \cite{cavalcante2022spatz} and Dustin \cite{ottavi2023dustin} address the Von Neumann bottleneck by reducing the impact of instruction fetching in multi-core architectures but at different scales: the former integrates a tightly coupled vector processing unit within each lightweight core, while the latter synchronizes all cores in lockstep by disabling individual instruction fetch units. To accommodate the intensive demands of AI workloads, HIPU \cite{zhao2023hipu} and SPEED \cite{wang2024speed} introduce dedicated matrix/tensor computation units. 

UVP falls short of achieving the maximum frequency due to two primary factors. First, each lane incorporates densely packed arithmetic units which introduce high fan-in logic, often implemented using dense and-or-invert (AOI) cells. This leads to significant local routing congestion. Second, the centralized EXE must maintain datapath connectivity with all surrounding lanes, resulting in placement contention and increased routing complexity.

To align with the execution unit configurations of Ara2 \cite{perotti2024ara2} and Yun \cite{perotti2023yun}, we employ a 16-lane setup with 32 vector registers for comparison. In terms of throughput, Ara2 \cite{perotti2024ara2} and Yun \cite{perotti2023yun} achieve at least 1.87$\times$ speedup in peak throughput under normalized conditions compared to our work. This advantage is likely due to their higher normalized frequency and better utilization across all execution units. However, our extension demonstrates superior area efficiency, outperforming these lane-based designs by 43.2\% and 30.9\%, respectively. This improvement arises from focusing on integer arithmetic computation and vector permutation while eliminating floating-point units, resulting in a more compact core area.

Our design is further compared with other micro-architectures. While Dustin \cite{ottavi2023dustin} achieves peak throughput with 2-bit quantization, our extension potentially surpasses multi-core architectures in terms of area and efficiency. This outcome stems from the fact that, despite efforts to equip scalar cores with vector capabilities, some datapaths within the pipeline remain idle during vector processing, resulting in significant silicon waste. However, a substantial gap exists between our design and AI-focused extensions, namely, SPEED \cite{wang2024speed} and HIPU \cite{zhao2023hipu}. The high throughput of these architectures is driven by the numerous PEs in their systolic arrays, optimized for tensor computations, and the minimal control overhead. Although the addition of execution units increases silicon area, the resulting area efficiencies still outperform our design due to the overwhelming throughput achieved by concurrently operating PEs, which effectively amortizes the silicon overhead. Nonetheless, it is worth noting that both SPEED and HIPU are specifically tailored for tensor computing in AI workloads and lack evaluation results in the domain of wireless signal processing.

\begin{figure}[!t]
  \centering
  \includegraphics[width=\linewidth]{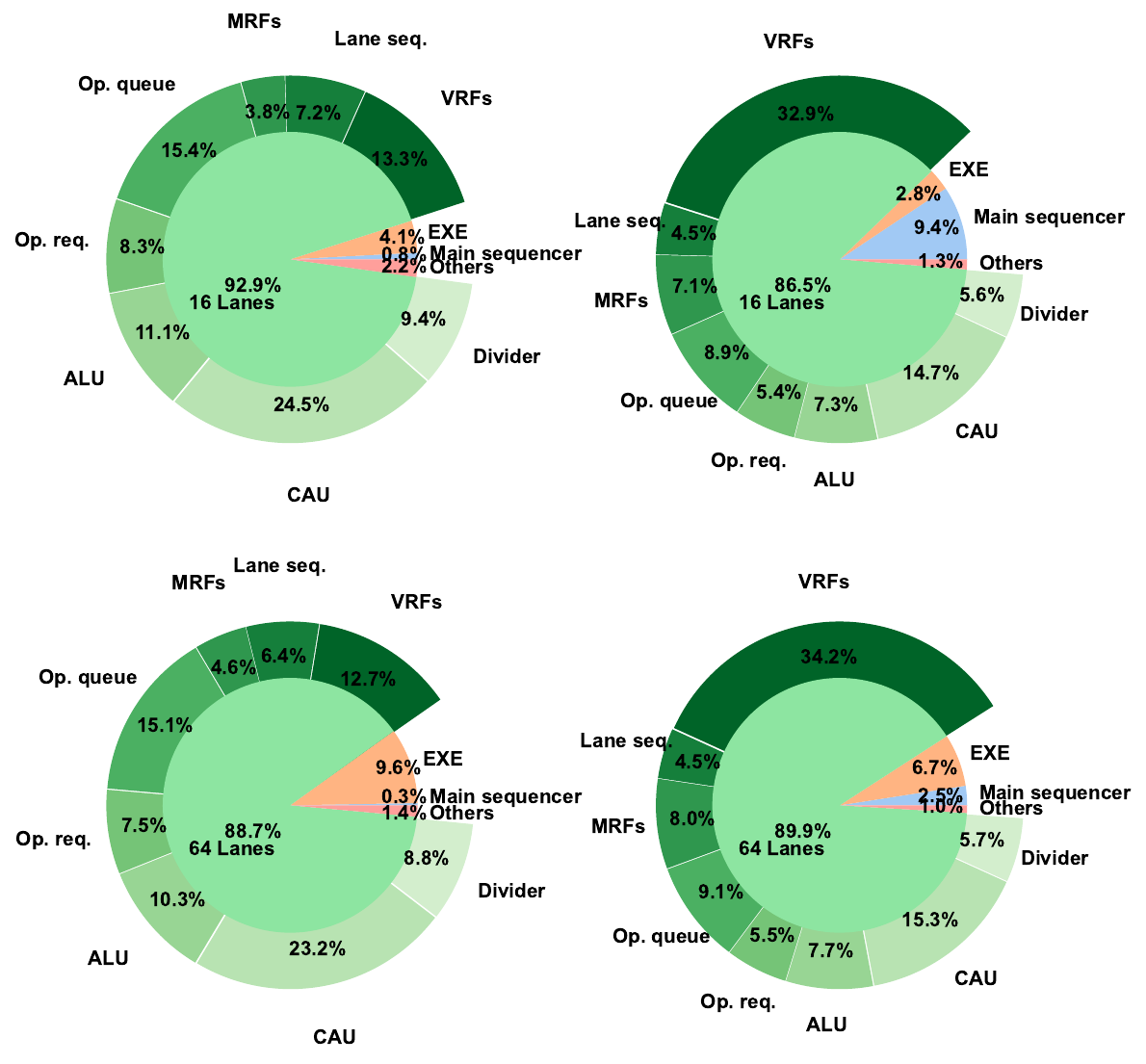}
  \vspace{-0.6cm}
  \caption{Area breakdown of vector extension under different configurations. Upper left: Lane$_{16}$Reg$_{32}$; Upper right: Lane$_{16}$Reg$_{1024}$; Bottom left: Lane$_{64}$Reg$_{32}$; Bottom right: Lane$_{64}$Reg$_{1024}$. The operand requester (\textit{Op. req.}) and operand queue (\textit{Op. queue}) within each lane handle operand fetching and temporary storage prior to the execution, respectively.}
  \label{fig:pie}
  \vspace{-0.6cm}
\end{figure}

\subsubsection{Area Breakdown}
\label{section:breakdown}
The area breakdown of the proposed extension across different configurations is illustrated in Fig. \ref{fig:pie}. Generally, the lanes occupy the largest area proportion, with minimal overhead from the main sequencer and EXE. When $r$ is small, the main sequencer, which includes the proposed UVP detection, has a negligible area share. However, as $r$ increases, the area contribution of the sequencer grows due to additional compare logic. Additionally, the number of PEs in EXE scales with $l$, resulting in a larger EXE area as the lane count increases.

The lane breakdown is depicted around the internal pie chart, revealing how the SRAM footprint significantly impacts the distribution. When $r$ is small, the computation logic -- including the ALU, CAU, and divider -- accounts for roughly 45\% of the lane area, while SRAMs contribute around 13\%. However, as $r$ increases, the computation logic proportion drops to 28\%, a shift that partly explains the decreased area efficiency at higher $r$ values. This insight suggests that adding more execution units within each lane could help improve area efficiency.

\begin{figure}[!t]
  \centering
  \includegraphics[width=\linewidth]{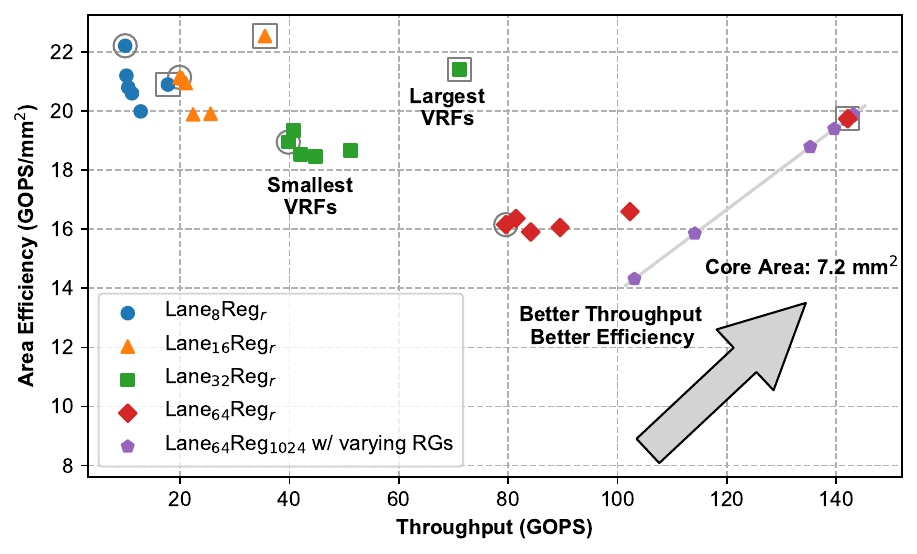}
  \vspace{-0.8cm}
  \caption{Comparsion of throughput and area efficiency under varying VRF and RG configurations. ($r$: Number of vector registers)}
  \label{fig:DSE}
  \vspace{-0.6cm}
\end{figure}

\subsubsection{Design Space Exploration}
To further assess the benefit of UVP, we examine register spilling and filling impacts. Our extension is tested with various lane counts and VRF SRAM configurations, with $l$ ranging from 8 to 64 and $r$ from 32 to 1024. Additionally, we set AVL to the maximum value $r=1024$ can support without spilling. Fig. \ref{fig:DSE} illustrates that as lane count rises, throughput increases, though area efficiency declines slightly due to the significant non-computation logic discussed in Section \ref{section:breakdown}. In the first four scatter plots, throughput generally improves as $r$ increases, with area efficiency remaining steady. Excluding $r=1024$, there is a 1.27$\times$ speedup when comparing the smallest VRFs (gray circles, $r=32$) and configurations with one spill ($r=512$). The largest VRFs (gray squares) deliver optimal throughput by avoiding register spilling and filling, achieving a 1.79$\times$ speedup.

To showcase the advantages of the large RGs, an additional evaluation is conducted with a fixed configuration, Lane$_{64}$Reg$_{1024}$, featuring a core area of 7.2 mm$^2$. The purple scatter plots in Fig. \ref{fig:DSE} illustrate the throughput increase as the size of the RG grows. Due to the reduced likelihood of data hazards and read/write conflicts , the maximum throughput is 1.38$\times$ higher than that of the $r=32$ configuration.

\section{Conclusion and Future Work}
\label{section:Conclusion}
We introduce Unlimited Vector Processing (UVP), a methodology that enables efficient and flexible vector processing for wireless baseband processing. UVP presents a novel architecture that supports non-power-of-two register grouping and hardware strip-mining. A programming model is proposed to seamlessly handle vectors of arbitrary lengths, significantly reducing strip-mining overhead and improving overall throughput. Additionally, UVP categorizes vector instructions into symmetric and asymmetric types, employing tailored load/store strategies that optimize execution across a wide range of workloads. The hardware implementation features hazard detection logic and optimized data pipelines, ensuring efficient execution of complex operations, particularly in fixed-point computations.

The experimental results demonstrate the effectiveness of UVP, achieving remarkable performance improvements with speedups of up to 3.0× in matrix multiplication and 2.1× in FFT kernels compared to traditional lane-based vector architectures. The synthesized RTL for a 16-lane configuration under SMIC 40nm technology occupies only 0.94 mm$^2$, delivering an impressive area efficiency of 21.2 GOPS/mm$^2$. 
Future work will focus on enhancing overall throughput by optimizing the EXE micro-architecture and incorporating multiple EXEs within the extension to mitigate pipeline stalling caused by limited execution units.

\bibliographystyle{ieeetr}

\bibliography{IEEEabrv,ref}

\begin{thebibliography}{10}

\bibitem{oliker2004scientific}
L.~Oliker, A.~Canning, J.~Carter, J.~Shalf, and S.~Ethier, ``Scientific computations on modern parallel vector systems,'' in {\em SC'04: Proceedings of the 2004 ACM/IEEE Conference on Supercomputing}, pp.~10--10, IEEE, 2004.

\bibitem{ferreira2023design}
L.~Ferreira, S.~Malkowsky, P.~Persson, S.~Karlsson, K.~{\AA}str{\"o}m, and L.~Liu, ``Design of an application-specific {VLIW} vector processor for {ORB} feature extraction,'' {\em Journal of Signal Processing Systems}, vol.~95, no.~7, pp.~863--875, 2023.

\bibitem{jouppi2023tpu}
N.~Jouppi, G.~Kurian, S.~Li, P.~Ma, R.~Nagarajan, L.~Nai, N.~Patil, S.~Subramanian, A.~Swing, B.~Towles, {\em et~al.}, ``Tpu v4: An optically reconfigurable supercomputer for machine learning with hardware support for embeddings,'' in {\em Proceedings of the 50th Annual International Symposium on Computer Architecture}, pp.~1--14, 2023.

\bibitem{git-rvv}
RISC-V, ``{riscv-v-spec},'' 2021.
\newblock [Online]. Available: \url{https://github.com/riscvarchive/riscv-v-spec}.

\bibitem{nannipieri2021risc}
P.~Nannipieri, S.~Di~Matteo, L.~Zulberti, F.~Albicocchi, S.~Saponara, and L.~Fanucci, ``A {RISC-V} post quantum cryptography instruction set extension for number theoretic transform to speed-up {CRYSTALS} algorithms,'' {\em IEEE Access}, vol.~9, pp.~150798--150808, 2021.

\bibitem{kuo2022risc}
Y.-M. Kuo, F.~Garc{\'\i}a-Herrero, O.~Ruano, and J.~A. Maestro, ``{RISC-V} galois field {ISA} extension for non-binary error-correction codes and classical and post-quantum cryptography,'' {\em IEEE Transactions on Computers}, vol.~72, no.~3, pp.~682--692, 2022.

\bibitem{gautschi2017near}
M.~Gautschi, P.~D. Schiavone, A.~Traber, I.~Loi, A.~Pullini, D.~Rossi, E.~Flamand, F.~K. G{\"u}rkaynak, and L.~Benini, ``Near-threshold {RISC-V} core with {DSP} extensions for scalable iot endpoint devices,'' {\em IEEE transactions on very large scale integration (VLSI) systems}, vol.~25, no.~10, pp.~2700--2713, 2017.

\bibitem{jiang2025hierarchical}
L.~Jiang, Y.~Shi, Y.~Liu, Q.~Deng, S.~Xu, Y.~Shen, F.~Ye, S.~Cao, and Z.~Jiang, ``A hierarchical dataflow-driven heterogeneous architecture for wireless baseband processing,'' in {\em Proceedings of the 30th Asia and South Pacific Design Automation Conference}, pp.~587--593, 2025.

\bibitem{spencer2004zero}
Q.~Spencer, A.~Swindlehurst, and M.~Haardt, ``Zero-forcing methods for downlink spatial multiplexing in multiuser mimo channels,'' {\em IEEE Transactions on Signal Processing}, vol.~52, no.~2, pp.~461--471, 2004.

\bibitem{tosato2002simplified}
F.~Tosato and P.~Bisaglia, ``Simplified soft-output demapper for binary interleaved {COFDM} with application to {HIPERLAN}/2,'' in {\em 2002 IEEE International Conference on Communications. Conference Proceedings. ICC 2002 (Cat. No. 02CH37333)}, vol.~2, pp.~664--668, IEEE, 2002.

\bibitem{behera2020efficient}
K.~C. Behera, ``An efficient low-latency algorithm and implementation for rate-matching and bit-interleaving in {5G NR},'' in {\em 2020 IEEE 3rd 5G World Forum (5GWF)}, pp.~565--571, IEEE, 2020.

\bibitem{janhunen2011fixed}
J.~Janhunen, T.~Pitkanen, O.~Silven, and M.~Juntti, ``Fixed-and floating-point processor comparison for mimo-ofdm detector,'' {\em IEEE Journal of Selected Topics in Signal Processing}, vol.~5, no.~8, pp.~1588--1598, 2011.

\bibitem{chen2020xuantie}
C.~Chen, X.~Xiang, C.~Liu, Y.~Shang, R.~Guo, D.~Liu, Y.~Lu, Z.~Hao, J.~Luo, Z.~Chen, {\em et~al.}, ``Xuantie-910: A commercial multi-core 12-stage pipeline out-of-order 64-bit high performance {RISC-V} processor with vector extension: Industrial product,'' in {\em 2020 ACM/IEEE 47th Annual International Symposium on Computer Architecture (ISCA)}, pp.~52--64, IEEE, 2020.

\bibitem{si5x280}
{{SiFive}}, ``{SiFive} intelligence {X280},'' 2022.
\newblock [Online]. Available: \url{https://sifive.cdn.prismic.io/sifive/70445cba-0549-475e-a538-5c09a402efbc_x280-datasheet-22G1.pdf}.

\bibitem{nx27v}
{Andes technology}, ``{AndesCore™ NX27V Processor},'' 2021.
\newblock [Online]. Available: \url{https://www.andestech.com/wp-content/uploads/AndesCore_NX27V_Product_Package_PB156_V1.2.pdf}.

\bibitem{semidynamic}
{Semidynamics}, ``{Semidynamics Vector Unit - Only 100\% customisable RISC-V Vector Unit},'' 2024.
\newblock [Online]. Available: \url{https://semidynamics.com/en/technology/vector-unit}.

\bibitem{cavalcante2019ara}
M.~Cavalcante, F.~Schuiki, F.~Zaruba, M.~Schaffner, and L.~Benini, ``Ara: A {1-GHz+} scalable and energy-efficient {RISC-V} vector processor with multiprecision floating-point support in 22-nm {FD-SOI},'' {\em IEEE Transactions on Very Large Scale Integration (VLSI) Systems}, vol.~28, no.~2, pp.~530--543, 2019.

\bibitem{perotti2022new}
M.~Perotti, M.~Cavalcante, N.~Wistoff, R.~Andri, L.~Cavigelli, and L.~Benini, ``A {`New Ara'} for vector computing: An open source highly efficient {RISC-V} {V} 1.0 vector processor design,'' in {\em 2022 IEEE 33rd International Conference on Application-specific Systems, Architectures and Processors (ASAP)}, pp.~43--51, IEEE, 2022.

\bibitem{perotti2024ara2}
M.~Perotti, M.~Cavalcante, R.~Andri, L.~Cavigelli, and L.~Benini, ``Ara2: Exploring single-and multi-core vector processing with an efficient {RVV} 1.0 compliant open-source processor,'' {\em IEEE Transactions on Computers}, 2024.

\bibitem{humblet2024msparq}
E.~Humblet, T.~Dupuis, Y.~Fournier, M.~H. AskariHemmat, F.~Leduc-Primeau, J.~P. David, and Y.~Savaria, ``{MSPARQ}: A {RISC-V} vector processor array optimized for low-resolution neural networks,'' in {\em 2024 IEEE 67th International Midwest Symposium on Circuits and Systems (MWSCAS)}, pp.~464--468, IEEE, 2024.

\bibitem{wang202430}
Y.~Wang, M.~Yang, C.-P. Lo, and J.~P. Kulkarni, ``30.6 {Vecim}: A 289.13 {GOPS/W RISC-V} vector co-processor with compute-in-memory vector register file for efficient high-performance computing,'' in {\em 2024 IEEE International Solid-State Circuits Conference (ISSCC)}, vol.~67, pp.~492--494, IEEE, 2024.

\bibitem{minervini2023vitruvius+}
F.~Minervini, O.~Palomar, O.~Unsal, E.~Reggiani, J.~Quiroga, J.~Marimon, C.~Rojas, R.~Figueras, A.~Ruiz, A.~Gonzalez, {\em et~al.}, ``Vitruvius+: An area-efficient {RISC-V} decoupled vector coprocessor for high performance computing applications,'' {\em ACM Transactions on Architecture and Code Optimization}, vol.~20, no.~2, pp.~1--25, 2023.

\bibitem{platzer2021vicuna}
M.~Platzer and P.~Puschner, ``Vicuna: A timing-predictable {RISC-V} vector coprocessor for scalable parallel computation,'' in {\em 33rd euromicro conference on real-time systems (ECRTS 2021)}, Schloss Dagstuhl-Leibniz-Zentrum f{\"u}r Informatik, 2021.

\bibitem{assir2021arrow}
I.~A. Assir, M.~E. Iskandarani, H.~R.~A. Sandid, and M.~A. Saghir, ``Arrow: A {RISC-V} vector accelerator for machine learning inference,'' {\em arXiv preprint arXiv:2107.07169}, 2021.

\bibitem{lee201445nm}
Y.~Lee, A.~Waterman, R.~Avizienis, H.~Cook, C.~Sun, V.~Stojanovi{\'c}, and K.~Asanovi{\'c}, ``A 45nm 1.3 {GHz} 16.7 double-precision {GFLOPS/W} {RISC-V} processor with vector accelerators,'' in {\em ESSCIRC 2014-40th European Solid State Circuits Conference (ESSCIRC)}, pp.~199--202, IEEE, 2014.

\bibitem{schmidt2021eight}
C.~Schmidt, J.~Wright, Z.~Wang, E.~Chang, A.~Ou, W.~Bae, S.~Huang, V.~Milovanovi{\'c}, A.~Flynn, B.~Richards, {\em et~al.}, ``An eight-core 1.44-{GHz} {RISC-V} vector processor in 16-nm {FinFET},'' {\em IEEE Journal of Solid-State Circuits}, vol.~57, no.~1, pp.~140--152, 2021.

\bibitem{johns2020minimal}
M.~Johns and T.~J. Kazmierski, ``A minimal {RISC-V} vector processor for embedded systems,'' in {\em 2020 Forum for Specification and Design Languages (FDL)}, pp.~1--4, IEEE, 2020.

\bibitem{ottavi2020mixed}
G.~Ottavi, A.~Garofalo, G.~Tagliavini, F.~Conti, L.~Benini, and D.~Rossi, ``A mixed-precision {RISC-V} processor for extreme-edge {DNN} inference,'' in {\em 2020 IEEE Computer Society Annual Symposium on VLSI (ISVLSI)}, pp.~512--517, IEEE, 2020.

\bibitem{wang2024speed}
C.~Wang, C.~Fang, X.~Wu, Z.~Wang, and J.~Lin, ``{SPEED}: A scalable {RISC-V} vector processor enabling efficient multiprecision {DNN} inference,'' {\em IEEE Transactions on Very Large Scale Integration (VLSI) Systems}, 2024.

\bibitem{li2022precision}
K.~Li, J.~Zhou, Y.~Wang, J.~Luo, Z.~Yang, S.~Yang, W.~Mao, M.~Huang, and H.~Yu, ``A precision-scalable energy-efficient bit-split-and-combination vector systolic accelerator for {NAS}-optimized {DNNs} on edge,'' in {\em 2022 Design, Automation \& Test in Europe Conference \& Exhibition (DATE)}, pp.~730--735, IEEE, 2022.

\bibitem{attari2022application}
M.~Attari, L.~Ferreira, L.~Liu, and S.~Malkowsky, ``An application specific vector processor for efficient massive {MIMO} processing,'' {\em IEEE Transactions on Circuits and Systems I: Regular Papers}, vol.~69, no.~9, pp.~3804--3815, 2022.

\bibitem{patsidis2020risc}
K.~Patsidis, C.~Nicopoulos, G.~C. Sirakoulis, and G.~Dimitrakopoulos, ``{RISC-V$^2$}: a scalable {RISC-V} vector processor,'' in {\em 2020 IEEE International Symposium on Circuits and Systems (ISCAS)}, pp.~1--5, IEEE, 2020.

\bibitem{lazo2022adaptable}
C.~R. Lazo, E.~Reggiani, C.~R. Morales, R.~F. Bagu{\'e}, L.~A.~V. Vargas, M.~A.~R. Salinas, M.~V. Cort{\'e}s, O.~S. {\"U}nsal, and A.~Cristal, ``Adaptable register file organization for vector processors,'' in {\em 2022 IEEE International Symposium on High-Performance Computer Architecture (HPCA)}, pp.~786--799, IEEE, 2022.

\bibitem{domingos2021unlimited}
J.~M. Domingos, N.~Neves, N.~Roma, and P.~Tom{\'a}s, ``Unlimited vector extension with data streaming support,'' in {\em 2021 ACM/IEEE 48th Annual International Symposium on Computer Architecture (ISCA)}, pp.~209--222, IEEE, 2021.

\bibitem{fernandes2024functional}
A.~Fernandes, L.~Crespo, N.~Neves, P.~Tom{\'a}s, N.~Roma, and G.~Falcao, ``Functional validation of the {RISC-V} unlimited vector extension,'' {\em IEEE Embedded Systems Letters}, 2024.

\bibitem{ta2022big}
T.~Ta, K.~Al-Hawaj, N.~Cebry, Y.~Ou, E.~Hall, C.~Golden, and C.~Batten, ``Big.{VLITTLE}: On-demand data-parallel acceleration for mobile systems on chip,'' in {\em 2022 55th IEEE/ACM International Symposium on Microarchitecture (MICRO)}, pp.~181--198, IEEE, 2022.

\bibitem{bedoukian2021software}
P.~Bedoukian, N.~Adit, E.~Peguero, and A.~Sampson, ``Software-defined vector processing on manycore fabrics,'' in {\em MICRO-54: 54th Annual IEEE/ACM International Symposium on Microarchitecture}, pp.~392--406, 2021.

\bibitem{ottavi2023dustin}
G.~Ottavi, A.~Garofalo, G.~Tagliavini, F.~Conti, A.~Di~Mauro, L.~Benini, and D.~Rossi, ``Dustin: A 16-cores parallel ultra-low-power cluster with 2b-to-32b fully flexible bit-precision and vector lockstep execution mode,'' {\em IEEE Transactions on Circuits and Systems I: Regular Papers}, vol.~70, no.~6, pp.~2450--2463, 2023.

\bibitem{mahurin2023qualocmm}
E.~Mahurin, ``Qualcomm{\textregistered} hexagon™ {NPU},'' in {\em 2023 IEEE Hot Chips 35 Symposium (HCS)}, pp.~1--19, IEEE Computer Society, 2023.

\bibitem{lin2007palf}
Y.-C. Lin, Y.-P. You, and J.-K. Lee, ``{PALF}: compiler supports for irregular register files in clustered {VLIW DSP} processors,'' {\em Concurrency and computation: practice and experience}, vol.~19, no.~18, pp.~2391--2406, 2007.

\bibitem{kuan2012compiler}
C.-B. Kuan and J.~K. Lee, ``Compiler supports for {VLIW DSP} processors with {SIMD} intrinsics,'' {\em Concurrency and Computation: Practice and Experience}, vol.~24, no.~5, pp.~517--532, 2012.

\bibitem{yu2025optimizing}
M.-S. Yu, H.-C. Chang, C.-T. Wang, Y.-W. Tien, T.-L. Chen, and J.-K. Lee, ``Optimizing computer vision algorithms with {TVM} on {VLIW} architecture based on {RVV},'' {\em The Journal of Supercomputing}, vol.~81, no.~1, p.~172, 2025.

\bibitem{codrescu2014hexagon}
L.~Codrescu, W.~Anderson, S.~Venkumanhanti, M.~Zeng, E.~Plondke, C.~Koob, A.~Ingle, C.~Tabony, and R.~Maule, ``Hexagon {DSP}: An architecture optimized for mobile multimedia and communications,'' {\em IEEE Micro}, vol.~34, no.~2, pp.~34--43, 2014.

\bibitem{ceva2020}
{CEVA Inc.}, ``{Introducing CEVA-XC16},'' 2020.
\newblock [Online]. Available: \url{https://www.ceva-dsp.com/wp-content/uploads/2020/03/CEVA-XC16_introduction_non_nda_public-v2.pdf}.

\bibitem{armmatmul}
{ARM}, ``{Learn the architecture - Optimizing C code with Neon intrinsics},'' 2025.
\newblock [Online]. Available: \url{https://developer.arm.com/documentation/102467/0201}.

\bibitem{git-ara}
Pulp-platform, ``{Ara Apps},'' 2021.
\newblock [Online]. Available: \url{https://github.com/pulp-platform/ara/tree/main/apps}.

\bibitem{acevedo20245g}
J.~Acevedo, F.~H. Fitzek, and P.~Seeling, ``{5G} channel estimation kernels on {RISC-V} vector digital signal processors,'' in {\em 2024 International Conference on Microelectronics (ICM)}, pp.~1--8, IEEE, 2024.

\bibitem{ScanGPU}
S.~Sengupta, M.~Harris, Y.~Zhang, and J.~D. Owens, ``{Scan Primitives for GPU Computing},'' in {\em SIGGRAPH/Eurographics Workshop on Graphics Hardware} (M.~Segal and T.~Aila, eds.), 2007.

\bibitem{lai2022efficient}
H.-M. Lai and J.-K. Lee, ``Efficient support of the scan vector model for {RISC-V} vector extension,'' in {\em Workshop Proceedings of the 51st International Conference on Parallel Processing}, pp.~1--8, 2022.

\bibitem{chang2019case}
C.-H. Chang, C.-C. Yang, J.-K. Lee, and Y.-C. Lin, ``Case study: Support {OpenCL} complex class for baseband computing,'' in {\em Proceedings of the International Workshop on {OpenCL}}, pp.~1--2, 2019.

\bibitem{bertuletti2023efficient}
M.~Bertuletti, Y.~Zhang, A.~Vanelli-Coralli, and L.~Benini, ``Efficient parallelization of {5G-PUSCH} on a scalable {RISC-V} many-core processor,'' in {\em 2023 Design, Automation \& Test in Europe Conference \& Exhibition (DATE)}, pp.~1--6, IEEE, 2023.

\bibitem{cavalcante2022spatz}
M.~Cavalcante, D.~W{\"u}thrich, M.~Perotti, S.~Riedel, and L.~Benini, ``Spatz: A compact vector processing unit for high-performance and energy-efficient shared-{L1} clusters,'' in {\em Proceedings of the 41st IEEE/ACM International Conference on Computer-Aided Design}, pp.~1--9, 2022.

\bibitem{perotti2023yun}
M.~Perotti, M.~Cavalcante, A.~Ottaviano, J.~Liu, and L.~Benini, ``Yun: An open-source, 64-bit {RISC-V}-based vector processor with multi-precision integer and floating-point support in 65-nm {CMOS},'' {\em IEEE Transactions on Circuits and Systems II: Express Briefs}, 2023.

\bibitem{zhao2023hipu}
W.~Zhao, G.~Yang, T.~Xia, F.~Chen, N.~Zheng, and P.~Ren, ``{HIPU}: A hybrid intelligent processing unit with fine-grained {ISA} for real-time deep neural network inference applications,'' {\em IEEE Transactions on Very Large Scale Integration (VLSI) Systems}, vol.~31, no.~12, pp.~1980--1993, 2023.

\end{thebibliography}


\end{document}